\documentclass[aps,pra,twocolumn,showpacs,superscriptaddress,nofootinbib]{revtex4-2}
\usepackage{graphicx}
\usepackage{amsmath}
\usepackage{amssymb}
\usepackage{epstopdf}
\usepackage{amsfonts}
\usepackage{dcolumn}
\usepackage{bigints}
\usepackage{hyperref}
\usepackage[normalem]{ulem}

\begin{document}

\begin{abstract}
    Highly-efficient quantum memories are essential for advancing quantum information processing technologies, including scalable quantum computing and quantum networks. We experimentally demonstrate a light storage and retrieval protocol in a tripod system using an ensemble of laser-cooled $^{87}$Rb atoms. The tripod system, which consists of three ground states and an excited state, offers rich dynamics: its use to coherently store and retrieve a weak probe pulse in the $^{87}$Rb $F=1$ ground state manifold leads to the interference of two spin-wave excitations during storage time that translate to an interference in the peak intensity of the retrieved probe pulse. Our work shows that these interferences, which manifest when varying the pulse sequence or energy level structure, can be controlled experimentally by varying the storage time, optical phase, and magnetic field strength. Theoretical simulations exhibit excellent agreement with the experimental results. This work demonstrates the rich dynamics and versatile capabilities of atomic tripod systems for light storage and retrieval, with key advantages over conventional $\Lambda$-systems, highlighting the potential of atomic tripod systems for applications in quantum information processing, quantum synchronization, and atomic memory protocols.
\end{abstract}

\title{Light Storage and Retrieval in an Atomic Tripod System }
\author{Shan Zhong}
\email{Shan.Zhong-1@ou.edu}
\address{Homer L. Dodge Department of Physics and Astronomy,
  The University of Oklahoma,
  440 W. Brooks Street,
  Norman,
Oklahoma 73019, USA}
\address{Center for Quantum Research and Technology,
  The University of Oklahoma,
  440 W. Brooks Street,
  Norman,
Oklahoma 73019, USA}
\author{A. J. Sudler}
\address{Homer L. Dodge Department of Physics and Astronomy,
  The University of Oklahoma,
  440 W. Brooks Street,
  Norman,
Oklahoma 73019, USA}
\address{Center for Quantum Research and Technology,
  The University of Oklahoma,
  440 W. Brooks Street,
  Norman,
Oklahoma 73019, USA}
\author{D. Blume}
\address{Homer L. Dodge Department of Physics and Astronomy,
  The University of Oklahoma,
  440 W. Brooks Street,
  Norman,
Oklahoma 73019, USA}
\address{Center for Quantum Research and Technology,
  The University of Oklahoma,
  440 W. Brooks Street,
  Norman,
Oklahoma 73019, USA}
\author{Alberto M. Marino}
\email{marino@ou.edu, marinoa@ornl.gov}
\address{Homer L. Dodge Department of Physics and Astronomy,
  The University of Oklahoma,
  440 W. Brooks Street,
  Norman,
Oklahoma 73019, USA}
\address{Center for Quantum Research and Technology,
  The University of Oklahoma,
  440 W. Brooks Street,
  Norman,
Oklahoma 73019, USA}
\address{Quantum Information Science Section, Computational Sciences and Engineering Division,
Oak Ridge National Laboratory, Oak Ridge, Tennessee 37831, USA}
\thanks{This manuscript has been authored in part by UT-Battelle, LLC, under
contract DE-AC05-00OR22725 with the US Department of Energy
(DOE). The publisher acknowledges the US government license
to provide public access under the DOE Public Access Plan
(http://energy.gov/downloads/doe-public-access-plan).}
\maketitle

\section{Introduction}

Quantum memory plays a central role in quantum information processing, acting as a temporary storage device for quantum states of light~\cite{RevModPhys.82.1041}. Developing quantum memory protocols that can efficiently and reliably store quantum information is essential for advancing quantum information processing technologies, including scalable quantum computation~\cite{PhysRevA.89.022317}, quantum networks~\cite{duan2001long, zhao2009millisecond, specht2011single}, and quantum repeaters~\cite{RevModPhys.83.33, PhysRevA.93.032327, PRXQuantum.2.040307}. Achieving this requires precise control and manipulation of light–matter interactions to enable efficient mapping, storage, and retrieval of photonic quantum states.

Among the available platforms, laser-cooled atomic ensembles stand out as ideal media for coherently transferring quantum information between light and atoms, owing to the atoms' long coherence time and high conversion efficiency~\cite{Cho:16, vernaz2018highly,RevModPhys.82.1041}. In particular, a tripod system---a four-level configuration that consists of three ground states and one excited state---offers unique opportunities for implementing complex quantum operations beyond standard $\Lambda$-type systems, such as
 interacting dark resonances, Hong-Ou-Mandel interference,
 and coupled dark-state polaritons (DSPs)~\cite{EPaspalakis_2002, PhysRevA.69.063802,PhysRevA.75.013810,PhysRevA.95.013818}.

Similar to the three-level $\Lambda$-configuration counterpart, the tripod configuration enables non-dissipative transfer of a probe field to  ground state atomic coherences via  two-photon transitions, with subsequent readout by retrieval control fields. The underlying dynamics can be effectively described using DSPs,  which are quasi-particles that have light and matter (atomic coherence) contributions~\cite{PhysRevA.65.022314, PhysRevLett.84.4232, PhysRevA.71.041801}. During the memory protocol, information carried by the light is coherently mapped to and subsequently retrieved from the atomic coherences. Importantly, the protocol preserves the information initially carried by the light. In contrast to the $\Lambda$-configuration, the tripod configuration supports the formation of two DSPs---as opposed to one DSP---within the ground-state manifold, allowing for richer dynamics, such as interference of the two DSPs as a result of phase imprinting during retrieval or relative phase accumulation between the DSPs during the storage time due to an external magnetic field-induced  energy level splitting within the ground state manifold.

The rich dynamics in a tripod system provide a promising platform for a range of applications in quantum technologies, enabling diverse quantum protocols such as optical storage and retrieval of light for a dual-channel quantum memory~\cite{PhysRevA.83.043815, PhysRevA.87.053830}, quantum squeezed state memory development~\cite{Losev_2016}, quantum gate operation~\cite{PhysRevA.70.032317, PhysRevA.75.062302}, and quantum synchronization in spin-1 systems~\cite{PhysRevLett.125.013601}. These protocols rely on the superposition of two atomic coherences or DSPs and are thus not possible in a $\Lambda$-configuration.

In this work, we experimentally demonstrate and numerically model light storage and retrieval in a tripod system using laser-cooled $^{87}$Rb atoms. By coherently storing and retrieving a weak probe pulse in the $^{87}$Rb $F=1$ ground state manifold, we observe an interference pattern that arises from the superposition of two DSPs and manifests itself in the efficiency or peak intensity of the retrieved probe. Specifically, our experimental results, supported by theoretical simulations, show that this interference pattern can be precisely controlled by varying the storage time, tuning the energy splitting of the ground-state sublevels (via the Zeeman effect), and adjusting the relative phase $\chi$ between the two control retrieval beams. Building on
previous work  such as Ref.~\cite{PhysRevA.83.043815}, we present a detailed experimental and theoretical analysis of the oscillations of the retrieval efficiency as a function of the magnetic field strength and the relative
phase $\chi$, revealing rich dynamics unique to the tripod configuration. Our results highlight the robustness and versatility of cold-atom tripod systems in light storage protocols and their potential as a building block in quantum information processing applications.

The remainder of this paper is organized as follows. Section~\ref{sec_experiment} introduces our experimental setup. Technical details on the apparatus and experimental sequence are relegated to Appendix~\ref{appendix:experiment}. Section~\ref{sec_results} presents our experimental results and direct comparisons with theory predictions and simulations. Theory background and simulation details are discussed in Appendix~\ref{appendix:theory}. Finally, Sec.~\ref{sec_conclusion} summarizes.

\section{Experimental Setup}\label{sec_experiment}

\begin{figure*}
    \centering
    \includegraphics[scale=.27]{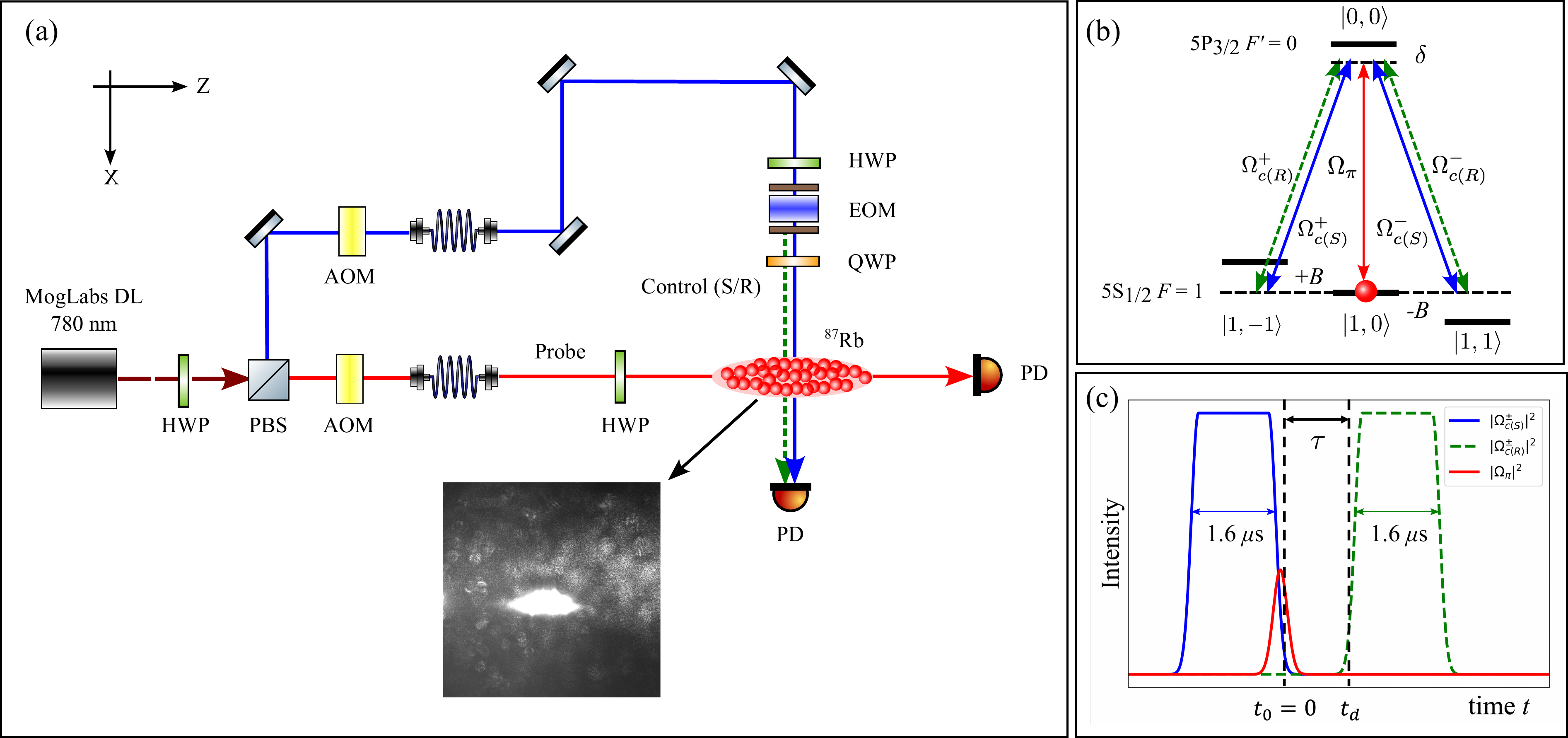}
    \caption{(a) Experimental layout. AOM: acousto-optic modulator, PBS: polarization beam splitter, HWP: half-wave plate, QWP: quarter-wave plate, PD: photodetector. The control storage beam (blue solid line) and the control retrieval beam (green  dashed line) are derived from the same laser, a MogLabs diode laser (DL). The relative phase $\chi$ between the $\Omega_{c(R)}^+$ and $\Omega_{c(R)}^-$ components of the control retrieval beam is controlled by an EOM and a QWP (see the text for details).  A magnetic bias field of strength $B_{\text{Gauss}}$ is applied along the $x$-axis. The inset at the bottom shows an absorption image of the atomic sample. MOT optics are omitted for clarity. (b) Energy level structure of tripod configuration. The four-level system consists of the three ground states in the $^{87}$Rb
    $F=1$ hyperfine ground state manifold and the $F'=0$ excited state ($D_2$ transition). The one-photon detuning $\delta$ is set to zero, while $B$ can be adjusted through the linear Zeeman shift by varying the magnetic field strength $B_{\text{Gauss}}$.  (c) Timing sequence of the control and probe beams for storage and retrieval. The times $t_0=0$ and $t_d$ are defined as the time when the intensity of the control storage pulse drops to $1/e^2$ of its maximum value and the time when the intensity of the control retrieval pulse rises to $1/e^2$ of its maximum value, respectively. The delay time is defined as $\tau = t_d-t_0$. The intensity of the probe pulse is much lower than that of the control pulse (the figure is not drawn to scale).}
    \label{fig:p1}
\end{figure*}

The tripod configuration is experimentally realized using four energy levels in a cold $^{87}$Rb atomic ensemble. As illustrated in Fig.~\ref{fig:p1}(b), we use
the $F=1$ ground state hyperfine manifold and the $F'=0$ excited state ($D_2$ transition), which form a closed system with no decay paths out of the four-level system. The experimental layout is illustrated in Fig.~\ref{fig:p1}(a). The ensemble of cold $^{87}$Rb atoms is held in an elongated magneto-optical trap (MOT). The atomic cloud has a length $L$ of approximately $3.0$~mm with a cross-sectional area of $A\approx\pi\times 0.66~ \mbox{mm}\times0.95~\mbox{mm}$
and contains $N\approx2.46(1) \times 10^{7}$ atoms. To prepare the atoms in the $F=1$,~$m_{F}=0$ state, we load the MOT and then apply polarization gradient cooling, which reduces the  atomic temperature to approximately 20~$\mu$K in the transverse direction (see Appendix~\ref{appendix:experiment} for details).  To minimize thermal effects, the temperature should be as low as possible. Subsequently, polarizing beams are applied to transfer as much of the population as possible from the $F=1$,~$m_{F}=\pm1$ states to the $F=1$,~$m_{F}=0$ state.

The probe and the control beams are resonant with the 5S$_{1/2}$ $F=1$ to 5P$_{3/2}$ $F'=0$ transition of the $D_2$ line, which has a wavelength of 780~nm; both beams are derived from the same diode laser. After passing through acousto-optic modulators (AOMs), both beams are coupled into polarization-maintaining single-mode optical fibers to ensure clean spatial modes. The probe beam propagates along the longitudinal axis of the MOT, where the optical depth (OD) is highest [the peak optical depth $\mbox{OD}_{\text{peak}}$ is approximately equal to $5.28(4)$]. The control beam propagates orthogonally to the probe beam to optimally couple the desired transitions. Importantly, owing to the orthogonal arrangement, the probe beam can be detected free from the strong ``background'' control light, enabling a high signal-to-noise ratio. After propagation through the MOT, two photodiodes  separately monitor the intensities of the probe beam and the control beam. Data are collected via an oscilloscope.

An external magnetic bias field of strength $B_{\text{Gauss}}$, which is used to lift the degeneracy of the $F=1$ ground state manifold through the linear Zeeman shift, is applied co-linearly with the control beam to set the quantization axis, i.e., the $x$-axis. Throughout the paper we report the resulting linear Zeeman shift $B$ between the $F=1$,~$m_{F}=0$ and $F=1$,~$m_{F}=\pm1$ states as a frequency ($2 \pi \times B$ corresponds to an angular frequency), which is related to the magnetic field strength $B_{\text{Gauss}}$ via $B=B_{\text{Gauss}}\times0.70$~MHz/G~\cite{steck-database}  (see also Appendix~\ref{appendix:experiment:bfield}).

To implement the storage and retrieval  of the probe field in the tripod system, we use a weak ($\approx 12$~$\mu$W) and small ($\approx 0.64$~mm 1/$e^2$ diameter) probe beam with $\pi$ polarization, and a comparatively  strong ($\approx 10$~mW) and large ($\approx 6.53$~mm 1/$e^2$ diameter) linearly polarized control beam. Due to the selection rules, the  $F=1$,~$m_{F}=0$~$\rightarrow$~$F'=0$,~$m_{F}=0$ transition can only be coupled by the $\pi$ polarized probe beam and not by the control beam. Since the control beam propagates along the quantization direction and is linearly polarized, the medium perceives it as a superposition of $\sigma^-$ and $\sigma^+$ polarizations, which couple the states $F=1$,~$m_{F}=\pm1$ and $F'=0$,~$m_{F}=0$, respectively, effectively acting as the required two control fields. As a result, the $F=1$,~$m_{F}=\pm1$ ground states are coupled to the $F=1$,~$m_{F}=0$ ground state via two-photon transitions, mediated by the probe and control fields. The relative phases $\chi_{S/R}$ of the storage ($S$)/retrieval ($R$) control beam,
defined as the phase difference between the $\sigma^+$ and $\sigma^-$ components of the control beam, can be controlled through the linear polarization angle $\theta_{S/R}$, which is measured with respect to an axis that lies in the $xy$-plane, according to $\chi_{S/R}=2\theta_{S/R}$. The relative phase $\chi_{S}$ of the control storage beam is arbitrarily set to zero, while the relative phase $\chi_R$ of the control retrieval beam can be adjusted ($\chi_R=\chi$ is constant during the retrieval). We experimentally control $\chi$ by tuning the polarization angle of the retrieval control beam $\theta_R$ using an electro-optic modulator (EOM) followed by a quarter-wave plate~\cite{Kaneshiro:16}. Since our EOM allows for a maximum polarization rotation of $\pi/2$, the largest achievable phase difference between the $\sigma^+$ and $\sigma^-$ components is
$\chi=\pi$.

The timing sequence of the control and probe beams is shown in Fig.~\ref{fig:p1}(c). The control storage pulse (Rabi coupling strength $\Omega_{c(S)}^\pm$) is first turned on for 1.6 $\mu$s to obtain a fully polarized state in $F=1$, $m_{F}=0$. This maximizes the optical depth for the $F=1$,~$m_{F}=0$~$\rightarrow$~$F'=0$,~$m_{F}=0$ transition and correspondingly optimizes the memory efficiency for storage of the probe beam~\cite{PhysRevA.90.055401}. As the control storage pulse falls off, a 300~ns short probe pulse (Rabi coupling strength $\Omega_\pi$) is applied. The intensities of the control storage and probe beams approach zero at approximately the same time [see Fig.~\ref{fig:p1}(c) and Table~\ref{tab1} for details]. When both the control and probe
pulses  are on, the probe field is coherently mapped to the collective atomic operators $\sigma_{12}$ and $\sigma_{23}$ in the $F=1$ ground state manifold. For notational convenience, we define the ground states as
$|1\rangle=|1,-1\rangle$,
$|2\rangle=|1,0\rangle$,
$|3\rangle=|1,+1\rangle$,
and the excited state as
$|4\rangle=|0,0\rangle$.
After a variable delay time $\tau$, the control retrieval pulse (Rabi coupling strength $\Omega_{c(R)}^\pm$), with a $1.6 ~\mu$s duration, is applied along the same direction as the control storage beam. The application of the control retrieval pulse  converts the probe field, which is stored  in the  atomic coherences during the delay time $\tau$, back into the optical field.
The delay time $\tau$ is, as indicated in Fig.~\ref{fig:p1}(c), defined as the time interval between the $1/e^2$ falling edge of the control storage pulse and the $1/e^2$ rising edge of the control retrieval pulse. We show below that the actual storage time deviates slightly from $\tau$. This is not surprising since the $1/e^2$-thresholds that define $\tau$ are chosen arbitrarily.

\section{Results and Analysis} \label{sec_results}

\begin{figure*}[t]
    \centering
    {\includegraphics[ scale=.45]{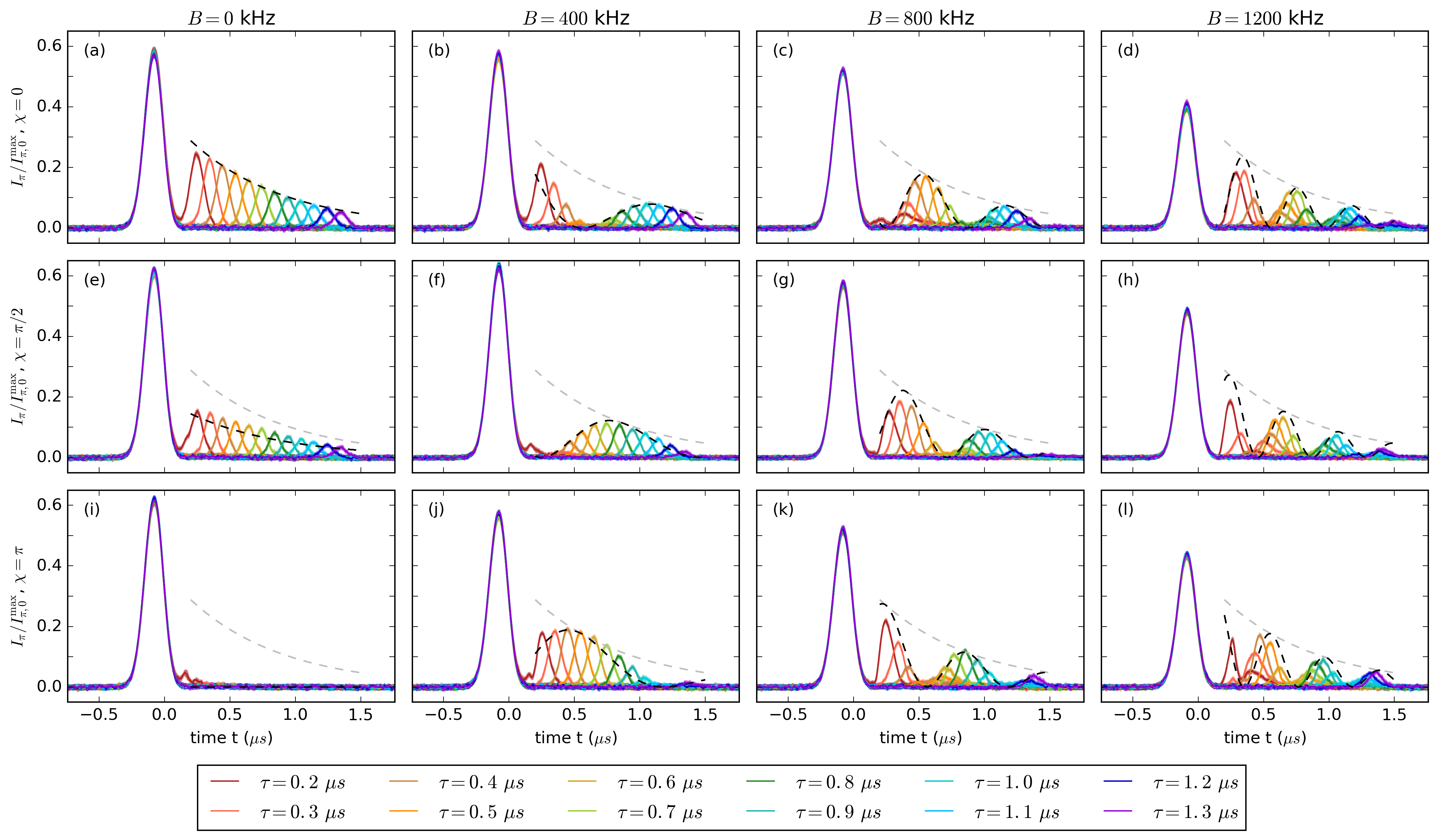}}
    \caption{Temporal profiles of the normalized probe intensity for different delay times $\tau$ (experimental data). The four columns show data for different Zeeman shifts $B$; specifically, $B=0$~kHz, $B=400$~kHz, $B=800$~kHz, and $B=1200$~kHz. The three rows show data for different relative phases $\chi$; specifically, $\chi=0$, $\chi=\pi/2$, and $\chi=\pi$. Each subplot shows the normalized probe pulses $I_{\pi}/I_{\pi,0}^{\text{max}}$ for various delay times $\tau$ ranging from 0.2~$\mu$s (dark red  solid line) to 1.3~$\mu$s (purple solid line); see the legend below the figure. All experimental data are the average of 10 independent runs. The shaded regions (very small) that surround each solid line represent the standard deviation of the mean. In (a), the black dashed line shows an exponential fit with rate $\gamma_c$ to the maxima of the normalized retrieved probe pulses; the fit yields $\gamma_c=2 \pi \times 0.111(3)$~MHz. For reference, this black dashed curve is replotted as a gray dashed curve in panels~(b)-(l). The black dashed lines in panels (b)-(l)
    are obtained by fitting the oscillations of the maxima of the normalized retrieved probe pulses to Eq.~(\ref{eq_envelope_dsp}), using  $\tau_0$ as a free parameter.}
    \label{fig:data1}
\end{figure*}

Figure~\ref{fig:data1} shows the experimentally measured intensity $I_{\pi}$ of the probe beam, recorded as a function of time at the end of the MOT, i.e., after the probe beam has passed through the atomic medium. The intensity $I_{\pi}$
is normalized to the peak intensity $I_{\pi,0}^{\text{max}}$ of the probe beam in the absence of the atomic medium. Figures~\ref{fig:data1}(a)-\ref{fig:data1}(l) each show experimental data for various delay times $\tau$. Specifically,  $\tau$ ranges from 0.2~$\mu$s (dark red solid line) to 1.3~$\mu$s (purple solid line), with increments of $0.1$~$\mu$s. The data shown in the four different columns are for four different Zeeman shifts $B$ while the data shown in the three different rows are for three different relative phases $\chi$. For fixed $\tau$, $B$, and $\chi$, the  normalized probe intensity $I_{\pi}/I_{\pi,0}^{\text{max}}$  contains two distinct contributions. The first contribution consists  of the highest  peak, whose  peak height is around $0.45-0.6$ and whose center is located at $t \approx -0.05$~$\mu$s. This peak depends comparatively weakly on  $\tau$, $B$, and $\chi$ and its shape is, to a good approximation,  Gaussian, i.e., it has  the same shape as the probe beam without the atomic medium. Correspondingly, we refer to this  peak as the transmitted probe signal, as it corresponds to the portion of the input probe beam that is not stored by the atomic medium, such that a maximum below $1$ indicates that a portion of the probe beam is absorbed by the medium.  The second contribution to $I_{\pi}/I_{\pi,0}^{\text{max}}$ corresponds to the signal that is seen for $t \gtrsim 0.2$~$\mu$s. This contribution, referred to as the retrieved probe  signal, depends comparatively strongly on $\tau$, $B$, and $\chi$. The retrieved probe  signal is the main focus of this paper.

Figure \ref{fig:data1}(a) shows the normalized probe intensity at a linear Zeeman shift of $B=0$~kHz and a relative phase of $\chi=0$. For $B=0$, the $F=1$ ground states of the system are degenerate,  and both the probe and control beams are resonantly coupled to their corresponding transitions. For $B=\chi=0$, the retrieval of the stored probe through application of the control retrieval beam should, in the absence of decoherence  mechanisms, be independent of the delay time $\tau$. The experimental data, however, show that the retrieved probe signal decreases  with increasing  delay time $\tau$. This indicates that there exist processes that degrade the atomic coherences between the states in the $F=1$ ground state manifold, thereby leading to a reduced retrieval efficiency with increasing $\tau$. Figure~\ref{fig:data1}(a) shows that the degradation depends exponentially on $\tau$.

Fitting the  maxima of the normalized retrieved probe pulses for different $\tau$ in Fig.~\ref{fig:data1}(a) to an exponential decay of the form $A e^{-2 \gamma_c \tau}$ ($\gamma_c$ denotes the decay rate), we find $\gamma_c=2\pi\times 0.111(3)$~MHz, which corresponds to a coherence time of $\tau_c= 1.43(4)$~$\mu$s.
The resulting fit is shown by the black dashed line. Note that the fitting function can alternatively be expressed as $A e^{2 \gamma_c \tau_0} e^{-2 \gamma_c (\tau+\tau_0)}$, which is useful for interpreting the results since it suggests that $\tau+\tau_0$ may be interpreted as an effective free evolution time or storage time  (see also below). To guide the eye, the fit is not only shown in  Fig.~\ref{fig:data1}(a) but replotted  in Figs.~\ref{fig:data1}(b)-\ref{fig:data1}(l)  using gray dashed lines. Since the signal  degradation is approximately exponential, the primary decoherence channel is likely related  to mechanisms such as coupling to the environment (e.g., via stray fields), magnetic spin wave dephasing, or generic field noise as opposed to  temperature-induced  dephasing, which would results in a Gaussian fall-off  of  the retrieved probe signal with increasing $\tau$~\cite{Saglamyurek_2021,PhysRevA.85.022318}. Our numerical simulations (see below) account for the decoherence through dissipative terms within a master equation framework. The simulations  use the $\gamma_c$ extracted from the experimental data as input and yield an exponential degradation of the retrieved probe, in agreement with the experiment.

Figures~\ref{fig:data1}(e) and~\ref{fig:data1}(i) show the normalized probe intensity for the relative phases $\chi = \pi/2$ and $\chi = \pi$, respectively [$B=0$ as in Fig.~\ref{fig:data1}(a)]. While the transmitted peak is essentially independent of $\chi$ (see above), the retrieved signal displays a clear $\chi$ dependence. At $\chi = \pi/2$, the amplitude of the retrieved probe is reduced compared to the $\chi=0$ case. Because the right- and left-handed circularly polarized portions of the control retrieval beam (Rabi coupling strengths $\Omega_{c(R)}^+$ and $\Omega_{c(R)}^-$, respectively) are no longer in phase, they imprint a different phase on the probe fields that are being read out from the two DSPs, resulting in partial destructive interference of the two
read-out fields. For $\chi=\pi$, the retrieval protocol is ineffective, i.e., a retrieved signal is not observed (full destructive interference). We emphasize that the relative phase is, in each panel,  the same for all delay times. As a consequence, we observe a reduction of the maxima of the normalized retrieved probe pulses for $B=0$ but no oscillations [at least not in the  representation chosen in Figs.~\ref{fig:data1}(a),  \ref{fig:data1}(e), and \ref{fig:data1}(i)]. To see oscillations, we consider finite $B$.

For finite magnetic field strengths, the degeneracy of the $F=1$ ground states is lifted and the control beams are no longer on resonance. The linear Zeeman splitting causes the two spin-wave components to acquire a relative phase of $\approx 2 \pi \times 2 B \tau$ during the storage stage, which leads to $\tau$-dependent oscillations of the maxima of the normalized retrieved probe
pulses. Figures~\ref{fig:data1}(b), \ref{fig:data1}(c), and \ref{fig:data1}(d) show the retrieved probe pulses for linear Zeeman shifts $B=400$, $800$, and $1200$~kHz, respectively. As $B$ increases, the linear Zeeman shift between the
ground-state sublevels grows, driving more rapid phase evolution between the two spin-wave components in the $F=1$ ground state manifold and thus faster
oscillations in the maxima of the normalized retrieved probe pulses in the representation chosen in Figs.~\ref{fig:data1}(b)-\ref{fig:data1}(d), \ref{fig:data1}(f)-\ref{fig:data1}(h), and \ref{fig:data1}(j)-\ref{fig:data1}(l). In the next paragraph, we use a simple DSP picture to describe the maxima of the normalized retrieved probe pulses [black dashed  lines in Fig.~\ref{fig:data1}(b)-\ref{fig:data1}(l)]. While the DSP picture nicely describes the observed behavior, it does not explain why the normalized intensity of the transmitted probe decreases---for fixed $\chi$---with increasing $B$. This behavior as well as the width of the retrieved peaks is, as we show below, reproduced by our numerical simulations.

The simplest DSP picture of the tripod system is based on treating the four-level system as being made up of two independent $\Lambda$-systems: the first $\Lambda$ system ($\Lambda_1$-system) consists of states $|1\rangle$, $|2\rangle$, and $|4\rangle$ while the second $\Lambda$ system ($\Lambda_2$-system) consists of states $|2\rangle$, $|3\rangle$, and $|4\rangle$~\cite{PhysRevA.71.041801, PhysRevLett.125.013601}. Within this picture, each of the two $\Lambda$ systems  supports one DSP. As in the usual $\Lambda$ system, the probe pulse transfers light to a collective atomic operator. Within the $\Lambda_1$ and $\Lambda_2$ systems, the light is transferred to $\sigma_{12}(z,t)$ and $\sigma_{23}(z,t)$, respectively. Assuming that the two DSPs evolve independently,  a relative phase of $2 \pi \times 2B \tau_{\text{eff}}$ is accumulated during the storage time. The factor of $2$ arises since the frequency splitting between the $|1\rangle$ and $|3\rangle$ states corresponds to an angular frequency of $2 \pi \times 2B$. The quantity $\tau_{\text{eff}}$ is interpreted as the effective free propagation time and referred to throughout this work as the actual storage time. While we expect  $\tau_{\text{eff}}$ to be approximately equal to $\tau$, we allow  $\tau_{\text{eff}}$ to be different from $\tau$. This is motivated by the fact that the maxima of the retrieval peaks in Fig.~\ref{fig:data1}(a) are not located exactly at $\tau$ but at slightly larger values: if  the times at which the retrieval signal takes its maximum are denoted by $t^*$, we find a linear relationship between $t^*$ and $\tau$ with a finite offset (i.e., the extrapolated $t^*$ value for $\tau=0$ is finite). Alternatively, we could say that the  definition of $\tau$ is not unique: $\tau$ could have, e.g., been defined through the decay or rise of the intensity to half  of the maximum as opposed to $1/e^2$ of the maximum.

In addition to the relative phase due to the free evolution, there is the externally imprinted relative phase $\chi$. Taken together, these two phases yield the total relative phase between the two DSPs, which gives rise to interference. Within the two independent $\Lambda$-systems picture, one obtains the following expression for the maxima of the normalized retrieved probe
pulses~\cite{PhysRevLett.125.013601},
\begin{eqnarray}
    \label{eq_envelope_dsp}\frac{I_{\pi}^{\text{peak}}(\tau)}{I_{\pi,0}^{\text{max}}}= \nonumber \\
    A \exp(-2 \gamma_c \tau)
    \cos ^2\left[ \frac{1}{2} \left(
    \chi + 2 \pi \times 2  B \left(\tau+\tau_0
    \right) \right)\right],
\end{eqnarray}
where $A$ and $\gamma_c$ are determined by a fit to the $B=\chi=0$ data (see above), the values of $B$ and $\chi$ are known from independent measurements, $\tau$ is set by the pulse sequence, and $\tau_0$ is treated as a fit parameter (one global fit parameter for all the data shown in Fig.~\ref{fig:data1}). With this, we have $\tau_{\text{eff}}=\tau+\tau_0$. Note that Ref.~\cite{PhysRevLett.125.013601} includes a Gaussian as opposed to an exponential factor in Eq.~(\ref{eq_envelope_dsp}); see above for  a brief discussion of this aspect. Excluding the data for $\tau = 0.2$~$\mu$s (for this short delay  time, the transmitted peak and the retrieved signal are not clearly separated in time), we find $\tau_0=0.065(1)$~$\mu$s. The results of the fit are shown, using $t$ instead of $\tau$, by the black dashed lines in  Figs.~\ref{fig:data1}(b)-\ref{fig:data1}(l). The overall agreement between the results predicted by the simple DSP picture and the experimental data is convincing. This indicates that the DSP picture captures key features of the dynamics when the probe is stored in the atomic degrees of freedom.

 \begin{figure*}[t]
    \includegraphics[width=0.9\linewidth]{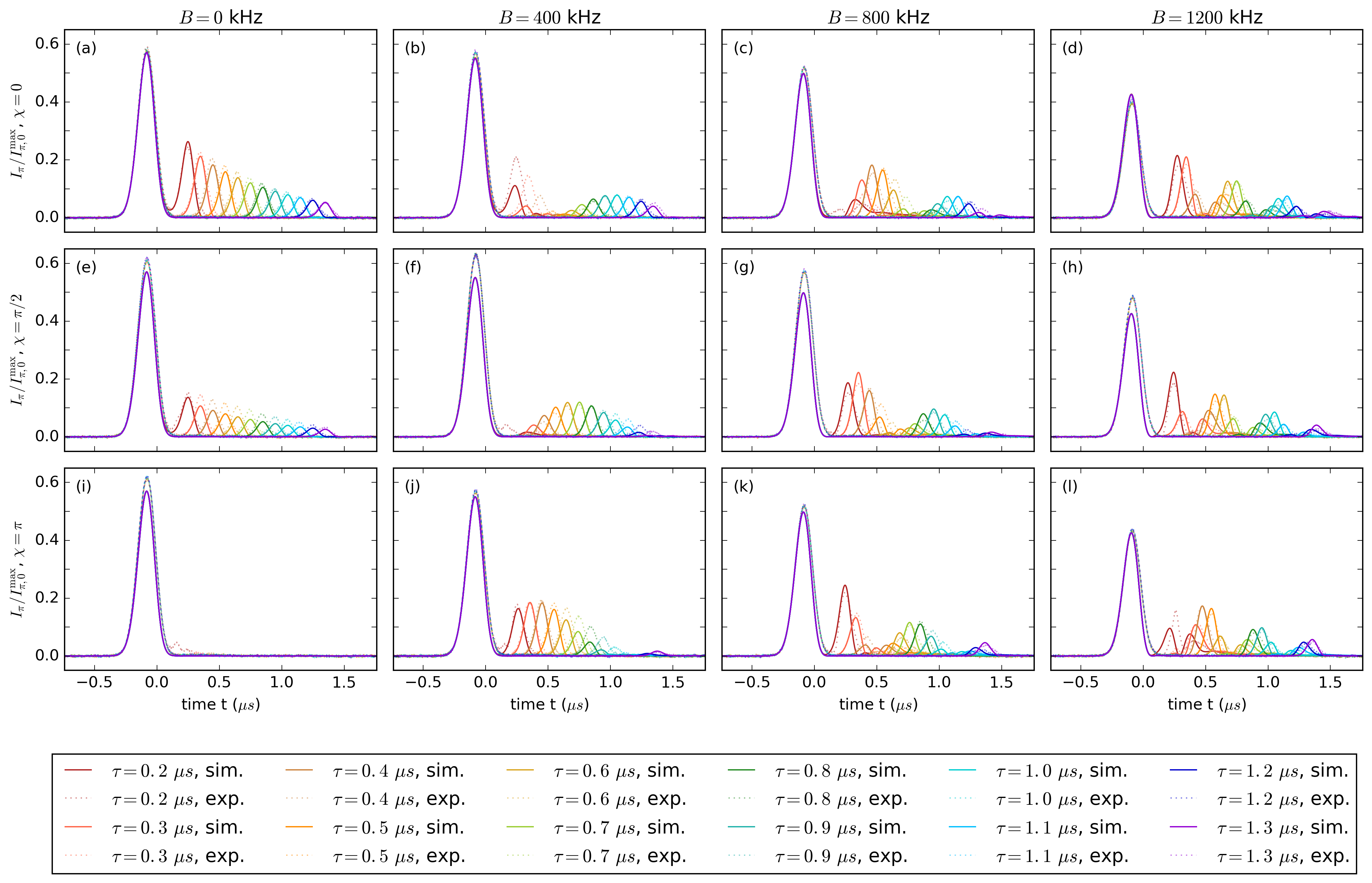}
    \caption{Comparison between the simulation results (thick solid lines) and the experimental data (thin dotted lines; same data as shown in Fig.~\ref{fig:data1}). The layout of the figure is the same as that of Fig.~\ref{fig:data1}. The simulation parameters are $\Omega_\pi^\text{max}=2\pi$~MHz, $\Omega_c^\text{max}= 2\pi \times 9.5$~MHz, $c\mu_a=2.118\times10^7$~$\mu$s$^{-2}$, and $\gamma_c=2\pi \times 0.111$ MHz.}
    \label{fig:slice_compare}
\end{figure*}

A more comprehensive theoretical description, which aims at capturing the entire time evolution---including the variation  of the input probe beam as it travels through the atomic medium---, is provided by a  model that is based on light propagating through a four-level atomic medium using the slowly-varying amplitude assumption. The evolution of the  probe field $\overline{\Omega}_\pi(z,t)$    is described by~\cite{PhysRevLett.125.013601,PhysRevA.71.041801,PhysRevA.99.043821}
\begin{align}
\label{eq_simulation_exact}
    \frac{1}{c}\frac{\partial\overline{\Omega}_\pi(z,t)}{\partial t}+\frac{\partial\overline{\Omega}_\pi(z,t)}{\partial z} = -\imath \frac{1}{2} \mu_a\sigma_{24}(z,t),
\end{align}
where $\overline{\Omega}_\pi(z,t)$ denotes the spatially-dependent Rabi frequency associated with the probe field (we consider a 1D MOT along the $z$-direction), $\mu_a$ the ``effective'' coupling constant for the  $|2\rangle \leftrightarrow |4\rangle$ transition ($\mu_a$ is proportional to the square of the atomic dipole moment), and $\sigma_{24}$  the spatially-dependent collective atomic operator for the $F=1$,~$m_{F}=0$ ground state to $F'=0$,~$m_{F}=0$ excited state transition (see Appendix~\ref{appendix:theory} for details). The time evolution of the full system---including all collective atomic operators $\sigma_{ij}(z,t)$ of the four-level system and the  optical field $\overline{\Omega}_\pi(z,t)$---is governed by a set of coupled Heisenberg-Langevin equations. These are detailed in Appendix~\ref{appendix:theoryframework}. The calibration of the input  Rabi couplings $\Omega_{\pi}^{\text{max}}$ and $\Omega_c^{\text{max}}$ as well as the value of $\mu_a$ through detailed comparisons of simulation data and experimental data is described in Appendix~\ref{appendix:calibration}.

Figure~\ref{fig:slice_compare}  compares the experimental data from Fig.~\ref{fig:data1} with our simulation results. While excellent overall agreement is observed, we note that the simulations do not accurately reproduce the retrieved signal for small $t$ ($t \approx 0.1$ to $0.2$~$\mu$s). We speculate that this might be due to finite occupation in states outside of the four levels treated in our simulations (see also Ref.~\cite{Laskar_2016}) or finite-temperature effects. If we use the theory results, as opposed to the experimental data (see above), to extract $\tau_0$ via a global one-parameter fit, we find $\tau_0=0.115(1)~\mu$s, which is a bit larger than the value found from fitting the experimental data [$\tau_0=0.065(1)$~$\mu$s, see above].  Encouragingly, though, the sign and magnitude of the $\tau_0$ values extracted from theory and experiment are comparable, implying that the extracted actual storage times $\tau_{\text{eff}}$ are rather similar.
The generally excellent  agreement between the experimental and simulation results is overall very encouraging, as it indicates that the simulations can be used to make quantitative predictions for other pulse sequences or to identify protocols that realize pre-specified target outcomes. The simulations also provide insights into the dynamics inside the MOT, which are not accessible experimentally. An example is shown in  Appendix~\ref{secII} (see Fig.~\ref{probe}).

\begin{figure*}[t]
    \centering
    {\includegraphics[scale=.46]{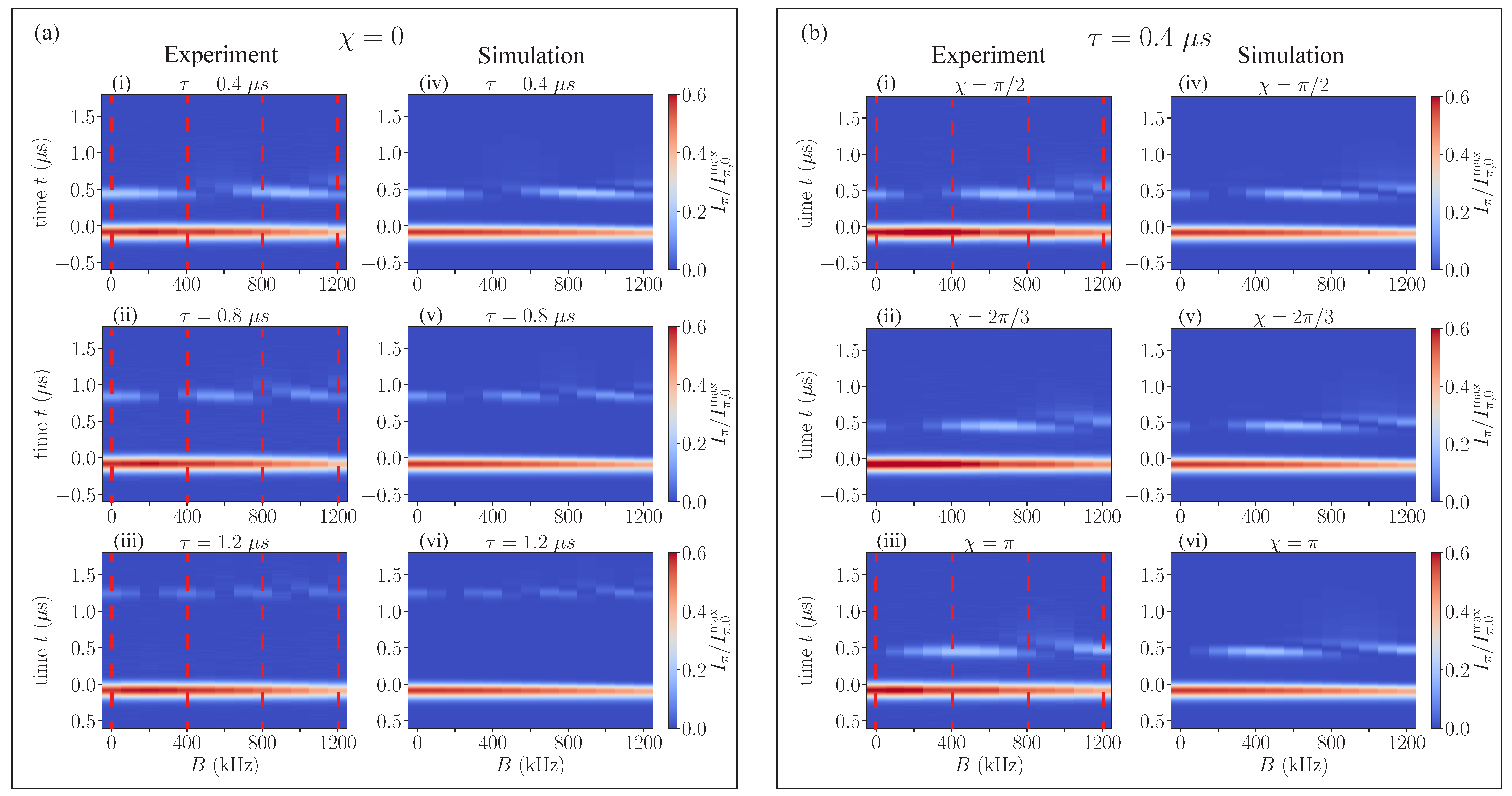}}
    \caption{Heatmap representation of the normalized probe intensity as a function of the linear Zeeman shift $B$ and  time $t$. The two  columns in subfigure (a) show experimental data and theoretical simulations, respectively, for  $\chi=0$ and varying delay time $\tau$: (ai) and (aiv) $\tau=0.4$ $\mu$s, (aii) and (av) $\tau=0.8$ $\mu$s, and (aiii) and (avi) $\tau=1.2$ $\mu$s. The two columns in subfigure (b) show experimental data and theoretical simulations, respectively,  for  $\tau=0.4$ $\mu$s and varying phase $\chi$: (bi) and (biv) $\chi=\pi/2$, (bii) and (bv) $\chi=2\pi/3$, (biii) and (bvi) $\chi=\pi$. The red dashed  lines in panels (ai), (aii), (aiii), (bi), and (biii) mark the linear Zeeman shifts corresponding to the time evolution traces of the retrieved probe shown in Fig.~\ref{fig:data1}. Theoretical simulations show excellent agreement with experimental data.}
    \label{fig:heatmap}
\end{figure*}

To better understand how the energy level splitting in the ground state manifold affects the probe retrieval dynamics, we scan the linear Zeeman shift
$B$ from 0 to 1200 kHz and plot the normalized intensity of the retrieved probe as a function of both magnetic field and time for various delay times $\tau$ and relative phases $\chi$. The experimental results (columns 1 and 3), along with our theoretical simulations (columns 2 and 4), are shown in Fig.~\ref{fig:heatmap}. It can be seen that the normalized intensity of the transmitted peak decreases with increasing $B$ field in the experimental data and the simulations.

The red dashed vertical lines in the first and third columns of Fig.~\ref{fig:heatmap} show the time evolution of the retrieved probe for $(B,\chi)$ combinations that are also shown in Figs.~\ref{fig:data1} and \ref{fig:slice_compare}. The first two columns of Fig.~\ref{fig:heatmap} display results for the relative phase $\chi=0$ for three different delay times $\tau$.
The experimental data in Figs.~\ref{fig:heatmap}(ai), \ref{fig:heatmap}(aii), and \ref{fig:heatmap}(aiii) show clear interference patterns of the normalized retrieved probe with changing linear Zeeman shift. As the delay time $\tau$ increases, the constructive and destructive interference alternates more frequently, reflecting the faster dynamical phase accumulation between the two collective atomic operators.

The third and fourth columns show results for $\tau=0.4$~$\mu$s and three different $\chi$. As $\chi$ increases, the positions of destructive interference shift---consistent with the dynamics described in Eq.~(\ref{eq_envelope_dsp})---
to lower linear Zeeman shifts. This shift occurs because the relative phase $\chi$ of the control retrieval beam is imprinted onto the optical field that is read out. Consequently, variations in $\chi$ allow one to modify at which linear Zeeman shifts or, equivalently, magnetic field strengths  destructive interferences occur. Notably, the positions of full destructive interference in
Figs.~\ref{fig:heatmap}(biii) and \ref{fig:heatmap}(bvi) align with those of
full constructive interference in Figs.~\ref{fig:heatmap}(ai) and \ref{fig:heatmap}(aiv), respectively. This correspondence confirms a relative phase shift of approximately $\pi$, consistent with the expected behavior from the two interfering DSPs. In closing we note that the simulations reproduce the experimentally observed interference patterns excellently, including both the periodic revival structures for fixed $\chi$ and the $\chi$-dependent shifts. This good agreement confirms the robustness of our theoretical framework in capturing the light-matter interaction dynamics of our tripod system.

\section{Conclusion}\label{sec_conclusion}

In summary, we have experimentally demonstrated and numerically validated the rich dynamics of the storage and retrieval of a probe beam in an atomic tripod system. Our results show that the retrieved probe signal is governed by the interference between the retrieved probe components,  which can be precisely controlled through the relative phase $\chi$ between the control retrieval beams, the  delay time $\tau$, and the applied magnetic field. Theoretical simulations based on a four-level system show excellent agreement with the experimental data, which confirms the rich dynamics and versatile capabilities for light storage and retrieval unique to the atomic tripod system. In contrast to conventional $\Lambda$ systems, the tripod system supports the simultaneous storage and interference of two DSPs, enabling richer manipulation of quantum states. This capability opens pathways to more advanced quantum protocols. Notably, the system could be suitable for storing two-mode squeezed states using dual DSPs~\cite{Losev_2016}, implementing quantum synchronization in spin-1 systems~\cite{PhysRevLett.125.013601}, and realizing slow-light soliton beam splitters~\cite{PhysRevA.99.043821}, all of which are foundational for quantum information processing and networking. Furthermore, the system’s sensitivity to external parameters such as magnetic field strength and optical phase suggests its potential for quantum sensing applications, including Hong-Ou-Mandel-type interferometry~\cite{PhysRevA.75.013810}.\\

\section*{Acknowledgment}
We thank Xylo Molenda and Tai Tran for insightful discussions. We thank the group of Grant Biedermann  for help with the design of the experimental apparatus and overall  support of the experiment. This work is supported by the W. M. Keck Foundation.  The computing for this project was in part performed at the OU Supercomputing Center for Education \& Research (OSCER) at the University of Oklahoma (OU). We are also grateful for the help and support provided by the Scientific Computing and Data Analysis section of Core Facilities at the Okinawa Institute of Science and Technology Graduate University (OIST).

\appendix

\section{Experimental Details}\label{appendix:experiment}

\subsection{Laser System Configuration}

\begin{figure*}
    \centering
    \includegraphics[scale=0.23]{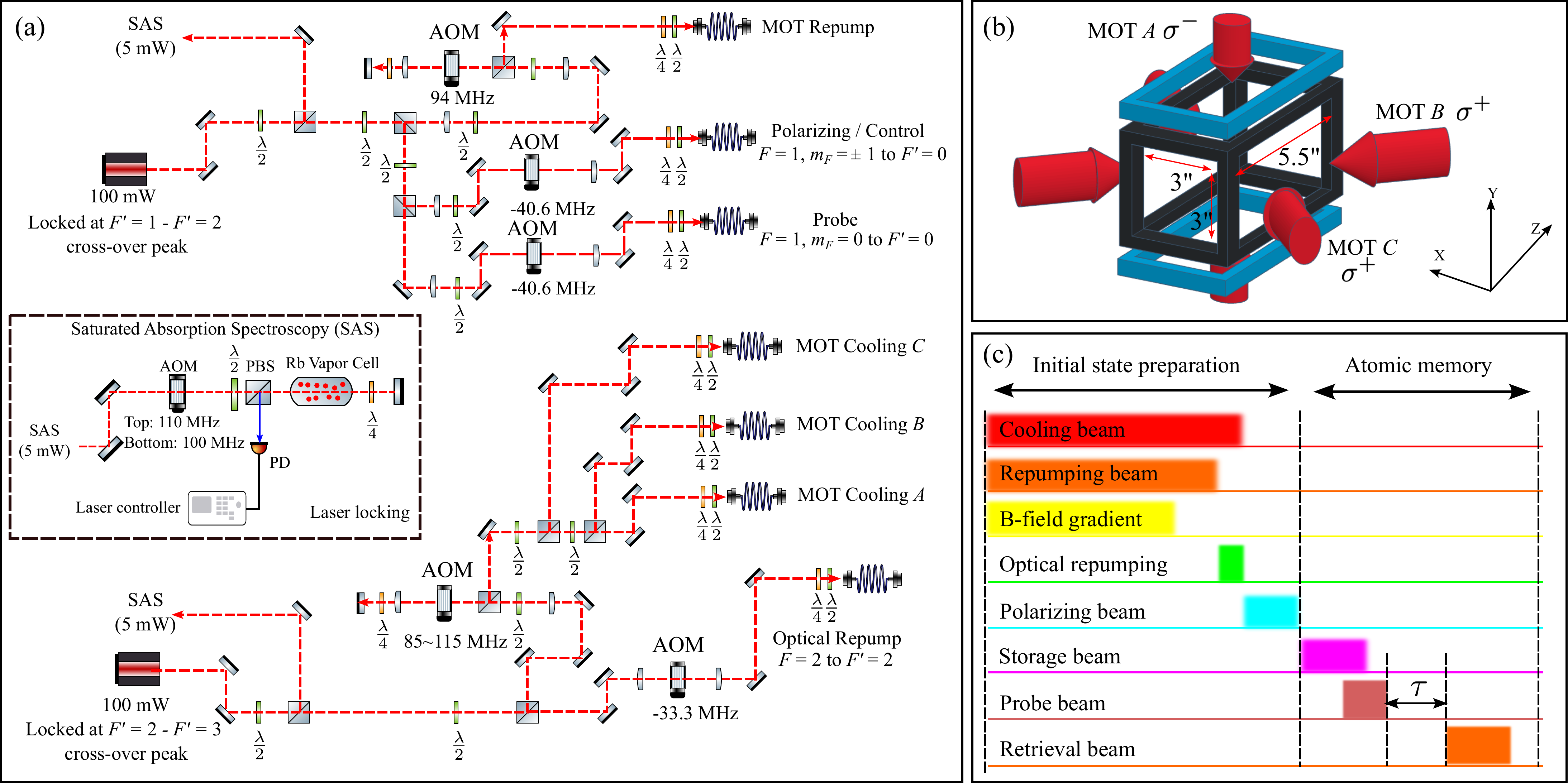}
    \caption{(a) Detailed schematic of the laser system. The diode laser shown in the bottom left generates the MOT cooling and optical pumping beams, while  the  diode laser shown in the top left generates the MOT repump, polarizing, probe, and control beams. A small portion of laser power ($\approx$ 5 mW) from each diode laser is directed to saturated absorption spectroscopy (SAS) setups for frequency stabilization, as shown in the inset. The AOM used in SAS for the locking of the laser shown in the top left has a central frequency of 110 MHz while the one used for the locking of the laser shown in the bottom left has a central frequency of 100~MHz. (b) Schematic of the MOT beam geometry and the MOT coil configuration. MOT beams $B$ and $C$ are $\sigma^+$ polarized and propagate orthogonally in the same horizontal plane at angles of 45$^\circ$ with respect to the $x$- and $z$-axes. MOT beam $A$ is $\sigma^-$ polarized and propagates vertically,  perpendicular to MOT beams $B$ and $C$. The magnetic coil frames have a rectangular shape with dimensions of 5.5" (long side) and 3" (short sides). A pair of anti-Helmholtz coils (blue frames) is used to generate the magnetic field gradient for MOT trapping. Three pairs of Helmholtz coils (black frames) are used to generate the bias magnetic field. (c) Timing sequence for one experimental cycle, which includes the initial state preparation and the atomic memory operation. The time axis is not drawn to scale.}
    \label{fig:sequence}
\end{figure*}

In this section, we detail the configuration of the optical fields used in our experiment, including the MOT cooling, MOT repump, optical repump, polarization, control, and probe beams. A detailed schematic of the laser setup is shown in Fig.~\ref{fig:sequence}(a). All the optical fields used in the experiment are generated from two Moglabs diode lasers, each emitting a linearly polarized, monochromatic beam at a wavelength of 780~nm and an output power of 100~mW.

The lasers are frequency-stabilized using saturated absorption spectroscopy (SAS), as illustrated in the inset of Fig.~\ref{fig:sequence}(a). The diode laser shown in the bottom left of Fig.~\ref{fig:sequence}(a) is locked to the crossover resonance between the $F'=2$ and $F'=3$ transitions of the $^{87}$Rb $D_2$ line and used to generate the MOT cooling and optical pumping beams.  The frequency and intensity of the MOT cooling beam are controlled via an AOM. During each experimental cycle, the frequency of the AOM is varied between 85 MHz and 115 MHz, allowing for a red detuning of up to 60 MHz from the $F=2 \rightarrow F'=3$ transition, which facilitates efficient atom loading and cooling. A double-pass AOM configuration is used to eliminate beam deflection resulting from frequency tuning. The optical repump beam, which is also derived from the laser that is shown in the bottom left of Fig.~\ref{fig:sequence}(a), is tuned with an AOM to the $F=2 \rightarrow F'=2$ transition and  used to  transfer residual atoms from the $F=2$ ground state to the $F=1$ ground state as part of the initial state preparation.

The diode laser shown in the top left of Fig.~\ref{fig:sequence}(a) is locked to the crossover resonance between the $F'=1$ and $F'=2$ transitions of the $D_2$ line, providing the MOT repump, polarization, probe, and control beams, which are tuned to their operational frequencies with AOMs. The MOT repump beam is resonant with the $F=1 \rightarrow F'=2$ transition of the $D_2$ line and serves to transfer atoms from the $F=1$ ground state back to the $F=2$ ground state, thereby returning them to the cooling cycle so that they can repeatedly interact with the MOT light.  The polarization, control, and probe beams are all tuned to the $F=1 \rightarrow F'=0$ transition. While the polarization and control beams originate from the same laser, they are used for different purposes: the polarization beam optically pumps atoms from the $F=1$, $m_F=\pm1$ states to the $F=1$, $m_F=0$ state as part of the initial state preparation, while the control beam is part of the memory protocol and facilitates two-photon transitions between the $F=1$ ground state sublevels mediated by the probe beam.

\subsection{Magneto-Optical Trap (MOT) Preparation}

The MOT is formed inside an ultra-high vacuum (UHV) scientific glass cell where the background pressure is maintained at $\approx 2.0 \times 10^{-10}$~Torr. The UHV environment minimizes decoherence caused by collisions between cold atoms and background gas particles, thereby preserving the coherence of the atoms in the ensemble.

The MOT cooling beam is split into three retro-reflected pairs (MOT~$A$ beams, MOT~$B$ beams, and  MOT~$C$ beams), which  form the standard six-beam MOT configuration, as illustrated in Fig.~\ref{fig:sequence}(b). The beam pair $A$, which is  aligned along the $y$-axis (direction of gravity), has $\sigma^-$ polarization. The remaining two beam pairs, which are  oriented in a plane orthogonal to the $y$-axis, have $\sigma^+$ polarization. The three orthogonal beam pairs intersect at the center of the vacuum cell to create a stable trapping region.

The magnetic field gradient, which is used for the creation of the MOT, is generated from a pair of anti-Helmholtz coils [blue coil frames in Fig.~\ref{fig:sequence}(b)], which are  driven by currents that are flowing in opposite directions. The rectangular geometry of the coil generates a field gradient that is stronger along the $x$ and $y$ directions than along the $z$ direction. As a result, the MOT is elongated along the $z$-axis, which is the direction of the probe beam. The elongation along $z$ enhances the OD, which is critical for strong light–atom interactions. In addition,  the MOT cooling beams along the $x$ and $z$ directions have a lower power (approximately 4 mW per beam) than the cooling beam along the $y$ direction (power of approximately 12 mW). This intentional imbalance in the power of the cooling beams further enhances the anisotropic shape of the MOT.

A separate set of three Helmholtz coil pairs [black coil frames in Fig.~\ref{fig:sequence}(b)] is used to generate a uniform magnetic field at the MOT location. Each pair of these coils carries current in the same direction and serves two main functions: canceling residual stray magnetic fields during the preparation stage and providing a controlled bias field for applying Zeeman shifts during the atomic memory operations.

\subsection{Timing sequence}

The timing sequence of the experiment is controlled by a commercial FPGA-based hardware system developed by M-Labs. This system provides nanosecond (4~ns) timing resolution and sub-microsecond latency, making it well-suited for our atomic memory protocol, which operates on the microsecond timescale. The timing sequence of the experimental cycle, including atom cooling, trapping, and the atomic memory process, is illustrated in Fig.~\ref{fig:sequence}(c).

During the MOT loading stage, the cooling beam is detuned by $-10$ MHz from the $F=2 \rightarrow F'=3$ transition, while the MOT repump beam is kept resonant with the $F=1 \rightarrow F'=2$ transition. Both beams are turned on for 4~s together with the magnetic field gradient to load atoms into the MOT. After 4~s, the MOT coils are switched off for 100~$\mu$s to allow eddy currents to decay. Next, the cooling beam is detuned by an additional 50~MHz for polarization gradient cooling (PGC), which lasts for 300~$\mu$s. The PGC  efficiently pumps the atoms from the $F=2$ ground state to the $F=1$ ground state to cool the atoms' temperature to the microKelvin regime. During the PGC, the MOT repump beam is turned off and a near-resonant optical repump beam, which couples the $F=2 \rightarrow F'=2$ transition, is turned on for 40~$\mu$s to transfer atoms efficiently into the $F=1$ ground state. Simultaneously, residual magnetic fields are canceled using three pairs of Helmholtz coils [see Fig.~\ref{fig:sequence}(b)]. After all the beams are turned off, a bias magnetic field is applied along the quantization direction ($x$-axis), and held for 40 $\mu$s to allow the field to stabilize before the atomic memory sequence, which is discussed in detail in the main text, is initiated.

\subsection{Bias Magnetic Field Calibration}\label{appendix:experiment:bfield}

The bias magnetic field is generated using three pairs of copper coils arranged in Helmholtz configurations. The bias magnetic field direction sets the quantization
axis during the light storage and retrieval processes. After state preparation—i.e., once most atoms have been optically pumped into the $F=1$, $m_F=0$ ground state—a bias magnetic field of strength $B_{\text{Gauss}}$ is applied by sending current through the coils.  This field lifts the degeneracy of the ground states in the $F=1$ manifold via the linear Zeeman effect, resulting in a frequency shift between the magnetic sublevels that is  described by
\begin{equation}
    B=g_F\mu_Bm_FB_{\text{Gauss}}/h,
\end{equation}
where $g_F$, $\mu_B$, and $h$  denote the Land\'e $g$-factor,  Bohr magneton, and Planck's constant, respectively. The $m_F=0$ state is unaffected by the linear Zeeman shift.

To calibrate the magnetic field strength, we use microwave spectroscopy. Our microwave source is a Hewlett-Packard HP 83732B synthesized signal generator, capable of driving transitions between hyperfine ground states. Specifically, we use it to couple the $F=1$, $m_F=0$ ground state to the $F=2$, $m_F=1$ ground state. The microwave signal is amplified using a radio frequency amplifier (ZVE-3W-83+) and emitted to the cold atom ensemble via a microwave horn antenna. After preparation of the cold atoms in the $F=1$,~$m_F=0$ state, we apply a 90~$\mu$s microwave pulse to drive the transition from the $F=1$,~$m_F=0$ state to the $F=2$,~$m_F=1$ state. The corresponding transition frequency, which we denote by $f_{0,1}$, is recorded. The transferred atoms are detected using absorption imaging on the $F=2 \rightarrow F'=3$ transition. The linear Zeeman shift is then determined by $B=f_{0,1}-f_{0,0}$, where $f_{0,0}$ is  the frequency of the clock transition between the $F=1$,~$m_F=0$ and $F=2$,~$m_F=0$ states.

We independently calibrate the magnetic field generated by each of the three coil pairs [black coil frames in Fig.~\ref{fig:sequence}(b)]. For each coil set, we scan the input current and measure the corresponding transition frequency to map out the relationship between the Zeeman shift $B$ and the current. This provides a precise calibration of the magnetic field strength as a function of the coil current. However, determining the exact zero-field condition from microwave spectroscopy alone is challenging, since the $B = 0$ condition also corresponds to the clock transition, which does not shift under the linear Zeeman effect. To resolve this ambiguity, we employ a magnetometer (Honeywell HMC2003), which converts the local magnetic field into an electric signal, allowing us to accurately identify the zero-field condition with $28$~Hz accuracy.

\subsection{Absorption Imaging for Atom Number and Temperature Measurement} \label{appendix:experiment:profile}

Absorption imaging is a powerful and widely used technique for quantitatively characterizing cold atomic clouds. It provides access to parameters such as the OD, the atom number, and the density distribution~\cite{ketterle1999}.

Our absorption imaging system consists of a CMOS camera (FLIR Blackfly S) and two lenses with focal length $f = 300$~mm, arranged in a 4$f$ configuration to achieve unit magnification. A collimated imaging beam, resonant with the $F=2 \rightarrow F'=3$ cycling transition of the $^{87}$Rb $D_2$ line, is aligned along the $z$-axis, which is the longitudinal axis of the MOT. The CMOS camera detects the attenuation of the imaging beam as it passes through the atomic cloud.

To perform absorption imaging, we first load cold atoms into the MOT. After the loading stage, all MOT-related beams and fields are turned off. A resonant imaging pulse is then applied for a duration of 50~$\mu$s to illuminate the atom cloud. While the imaging pulse is on, we additionally apply a weak repump beam, which is resonant with the $F=1 \rightarrow F'=2$ transition and moves atoms that have decayed to the $F=1$ ground state back to the $F=2$ ground state. This
ensures that the number of atoms that absorb the imaging light and, correspondingly,  contribute to the measured signal remains constant during the imaging process.

Three images are recorded to extract the optical depth: one with atoms and the imaging beam present ($I_a$), one with only the imaging beam present ($I_0$), and one background image with neither atoms nor imaging beam present ($I_{\text{bg}}$). The OD is calculated pixel-by-pixel using the Beer–Lambert law:
\begin{equation}
    \text{OD}(x,y)=-\ln\frac{I_a(x,y)-I_{\text{bg}}(x,y)}{I_0(x,y)-I_{\text{bg}}(x,y)}.
\end{equation}
The total atom number $N$ is obtained by integrating  $\text{OD}(x,y)$  over the entire image and dividing by the effective absorption cross section $\sigma_{\text{eff}}$,
\begin{equation}
    N=\frac{1}{\sigma_{\text{eff}}}\iint \text{OD}(x,y)dxdy,
\end{equation}
where $\sigma_{\text{eff}}=\sigma_0(1+I_{\text{imag}}/I_{\text{sat}})^{-1}$ with $\sigma_0$, $I_{\text{sat}}$, and $I_{\text{imag}}$ denoting the resonant cross section, saturation intensity of the $F=2 \rightarrow F'=3$ imaging transition, and intensity of the imaging beam, respectively (the ratio $I_{\text{imag}}/I_{\text{sat}}$ is approximately equal to $0.62$). This method provides a reliable
way to determine the atom number in the cloud and to extract other relevant parameters critical to the analysis of light–matter interactions  such as the size and temperature of the atomic cloud.

The transverse temperature of the cold atomic cloud is measured using the time-of-flight (TOF) method. After the trapping fields are turned off, the atoms expand freely for a few milliseconds before the imaging beam is applied. During this time, the velocity distribution of the atoms, governed by the Maxwell-Boltzmann distribution, translates into a spatial distribution due to the absence of confining forces.  The expansion of the cloud can therefore  be related to the transverse temperature $T$ of the cloud via~\cite{TomaszMBrzozowski_2002}
\begin{equation}
    \lambda_{x/y}(t)=\sqrt{\lambda_{0,x/y}^2+\frac{k_BT}{m}t^2},
    \label{eq.temp}
\end{equation}
where $\lambda_{0,x}$ and $\lambda_{0,y}$ are the initial widths of the cloud in the $x$ and $y$ directions, respectively, $\lambda_{x/y}(t)$
are the widths after a free expansion time $t$, and $m$ is the mass of the $^{87}$Rb atom ($m \approx 1.443 \times 10^{-25}$~kg)~\cite{steck-database}. The initial widths $\lambda_{0,x/y}$ and the time-dependent widths $\lambda_{x/y}(t)$
are determined experimentally using absorption imaging. Each image requires a full experimental cycle. By taking a series of images at different TOF delays, we extract the cloud widths $\lambda_x(t)$ and $\lambda_y(t)$ in the $x$ and $y$ directions, respectively, by fitting each image with a two-dimensional Gaussian profile. These measurements are then fitted to Eq.~(\ref{eq.temp}) to extract the temperature along each axis. The  reported  transverse temperature $T$ of $20$~$\mu$K is the average of the temperatures along the $x$ and $y$ axis.

\section{Theory Details}\label{appendix:theory}

\subsection{Theoretical Framework}\label{appendix:theoryframework}

Our theoretical framework treats $N$ non-interacting four-level atoms, i.e., we account  for the $F=1$ ground state hyperfine manifold of the $^{87}$Rb atom
and the $F'=0$ excited state ($D_2$ transition). We choose our energy scale such that $E_2$ has an energy of $0$ ($E_j$ is the energy of state $|j\rangle$). The energies $E_j$ are expressed in terms of angular frequencies:
$E_1=2 \pi \times B$ and $E_3=-2 \pi \times B$.

Throughout, we work on resonance, i.e., the frequency $\omega_{\pi}$ of the field that couples states $|2\rangle$ and $|4\rangle$ [Rabi coupling strength $\Omega_{\pi}(t)$] as well as the frequency $\omega_c$ of the field that couples states $|1\rangle$ and $|4\rangle$ and states $|3\rangle$ and $|4\rangle$ [Rabi coupling strength $\Omega_c(t)$] are set equal to the frequency of the $|2\rangle \leftrightarrow |4\rangle$ transition (approximately $2 \pi \times 384$~THz).
Even though the $|1\rangle \leftrightarrow |4\rangle$ and $|3\rangle \leftrightarrow |4\rangle$ transitions are both coupled by fields that are derived from the same laser (see the main text and Appendix~\ref{appendix:experiment}), we define a Rabi coupling strength $\Omega_c^+(t)$ that drives the  $|1\rangle \leftrightarrow |4\rangle$ transition and a Rabi coupling strength $\Omega_c^-(t)$ that drives the  $|3\rangle \leftrightarrow |4\rangle$ transition.
The experimental set-up enforces $|\Omega_c^+(t)|=|\Omega_c^-(t)|$ but allows for a relative phase between the two Rabi coupling strengths. To account for this phase difference, we write $\Omega_c^{\pm}(t)=\exp[\imath \chi^{\pm}(t)] |\Omega_c^{\pm}(t)|$. Working within the dipole approximation and making the rotating wave approximation, the Hamiltonian $H$ in the rotating frame reads
\begin{equation}
    \renewcommand*{\arraystretch}{1.5}
    H = \begin{pmatrix}
            E_1 & 0 & 0   & -\frac{\Omega_c^{+}}{2} \\
            0   & 0 & 0   & -\frac{\Omega_{\pi}}{2} \\
            0   & 0 & E_3 & -\frac{\Omega_c^{-}}{2} \\
            -\frac{(\Omega_c^{+})^*}{2} & -\frac{(\Omega_{\pi})^*}{2}
                                    & -\frac{(\Omega_c^{-})^*}{2}
                                    & E_4
        \end{pmatrix}.
\end{equation}

\begin{figure}
    \centering
    \includegraphics[width=0.85\linewidth]{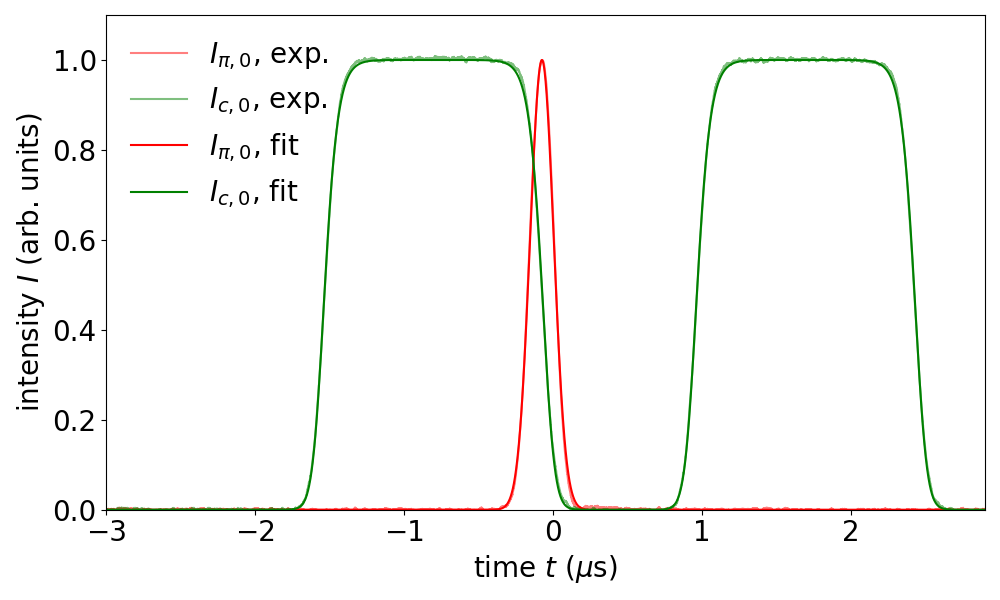}
    \caption{Comparison of experimentally measured intensities and our fits for
    $\tau=0.4$~$\mu$s in the absence of the atomic medium. Faint green and faint red lines show, respectively, the experimentally measured intensity $I_{\pi,0}$ of the probe beam and the experimentally measured intensity $I_{c,0}$ of the control beam as a function of time (the approximately constant background has been subtracted). The intensities are shown in arbitrary units (see text for details). The experimental data are averaged over 10 runs. The darker green and red lines show the results of our fits.
    The agreement between the fit and the experimental data is excellent (in fact, the agreement is so good that it is challenging to discern the two curves).
    The time $t_0=0$ is chosen such that the intensity of the control  storage beam  has decreased to $1/e^2$ of its maximum.}
    \label{fig1}
\end{figure}

The functional form  of $|\Omega_c^{\pm}(t)|$ and $|\Omega_{\pi}(t)|$ is, except for overall factors that we denote by $\Omega_c^{\max}$ and $\Omega_{\pi}^{\max}$, set by the square-root of the experimentally measured intensities $I_{c,0}(t)$ and $I_{\pi,0}(t)$, measured in the absence of any atoms. Throughout, the time $t$ at which the intensity of the storage pulse during turn-off has decayed to $1/e^2$ of its maximum value, is denoted by $t_0$; without loss of generality, we set $t_0=0$. As in the main text, the time  at which the intensity of the retrieval
pulse during turn-on  has reached $1/e^2$ of its maximum value is denoted by
$t_d$ and we use $\tau=t_d-t_0$. The faint green and red lines in Fig.~\ref{fig1} show the experimentally measured $I_{c,0}(t)$ and $I_{\pi,0}(t)$, respectively
(after subtracting the approximately constant background signal  and employing arbitrary units on the $y$-axis). We use the following parametrization for $|\Omega_c^{\pm}(t)|^2$ and $|\Omega_{\pi}(t)|^2$:
\begin{equation}
\label{eq_beam_pm}
    |\Omega_c^{\pm}(t)|^2 =\left[ \Omega_{c}^{\max}
    \sum_{l=1}^4\frac{(-1)^{l-1}}{2} \tanh \left(\frac{t-t_l}{\tau_l} \right) \right]^2
\end{equation}
and
\begin{equation}
\label{eq_beam_pi}
    |\Omega_{\pi}(t)|^2 = \left\{\Omega_{\pi}^{\max}
    \exp \left[-\frac{1}{2}\frac{(t-t_{\pi})^2}{\tau_{\pi}^2}\right] \right\}^2.
\end{equation}
In Eq.~(\ref{eq_beam_pm}), the $l=1$ and $2$ terms describe the storage part of the control beam  while the $l=3$ and $4$ terms describe the retrieval part of the control beam [the main text uses $\Omega_{c(S)}^{\pm}(t)$ and $\Omega_{c(R)}^{\pm}(t)$ to distinguish the storage and retrieval beams, respectively]. To determine the times $t_l$ and $t_{\pi}$ and the time constants $\tau_l$ and $\tau_{\pi}$, we fit Eqs.~(\ref{eq_beam_pm}) and (\ref{eq_beam_pi}) to the experimental data shown in Fig.~\ref{fig1} for $\tau=0.4$~$\mu$s. The thick green and red lines show the
resulting fits. The  fitting parameters $t_l$, $\tau_l$, $t_{\pi}$, and $tau_{\pi}$, which are independent of $\tau$, are listed in Table~\ref{tab1}. The agreement between the two fits and the experimental data is excellent. The values of $\Omega_{c}^{\max}$ and $\Omega_{\pi}^{\max}$ cannot be determined through these fits since it is challenging to obtain a precise calibration of the conversion from power to Rabi coupling strength with our experimental setup. We thus instead adjust the values of $\Omega_{c}^{\max}$ and $\Omega_{\pi}^{\max}$ to match experimental data that are taken in the presence of atoms (see Appendix~\ref{appendix:calibration} for details). In our treatment of the relative phase $\chi$, we add the phase exclusively to the coupling between states $|3\rangle$ and $|4\rangle$ during the retrieval; i.e., $\Omega_c^-(t)=|\Omega_c^-(t)|$ during the storage while $\Omega_c^-(t)=|\Omega_c^-(t)|\exp(\imath \chi)$ during the retrieval.

\begin{table}[ht]
    \centering
    \begin{tabular}[t]{cc}
        Parameter & Value  \\ \hline
        $\tau_1$ & $0.1017(3)$~$\mu\text{s}$ \\
        $\tau_2$ & $0.1022(3)$~$\mu\text{s}$ \\
        $\tau_3$ & $0.1021(3)$~$\mu\text{s}$ \\
        $\tau_4$ & $0.1024(3)$~$\mu\text{s}$ \\
        $t_1$ & $-1.577(2)$~$\mu$s \\
        $t_2$ & $-0.031(2)$~$\mu$s \\
        $t_3$ &
        $0.024(2)~\mu\mbox{s} + \tau$
        \\
        $t_4$ &
        $1.568(1)~\mu\mbox{s} + \tau$
         \\
        $t_{\pi}$ & $-0.07541(6)$~$\mu$s \\
        $\tau_{\pi}$ & $0.11167(7)$~$\mu$s \\
     \end{tabular}
    \label{tab1}
    \caption{Fitting parameters and their uncertainties [see Eqs.~(\ref{eq_beam_pm}) and (\ref{eq_beam_pi})]. The times $t_3$ and $t_4$
    depend on the delay time $\tau$, which is varied ($\tau$ is known from experiment).}
\end{table}

Treating the atomic spin degrees of freedom via a set of coupled Heisenberg-Langevin equations~\cite{PhysRevLett.125.013601}, our theory framework accounts for two dissipative processes: the first originates from the finite lifetime of state $|4\rangle$ and the second from the coupling to the environment. The excited state $|4 \rangle$ has a lifetime $\tau_{\text{life}}$ of $26.24(4)$~ns, which corresponds to a decay rate of $\Gamma= (\tau_{\text{life}})^{-1}=2 \pi \times 6.065(9)$~MHz. The excited state decays with equal probability to states $|1\rangle$, $|2\rangle$, and $|3\rangle$, i.e., the branching ratio is $1/3$. We model the decay via three Lindbladian ${\cal{L}}_j$, one for each of the three decay paths: ${\cal{L}}_j=\sqrt{\Gamma/3} |j\rangle \langle 4|$, where $j=1-3$.
To account for all other decoherence mechanisms, we introduce a phenomenological dissipative term that is characterized by the rate $\gamma_c$ whose value is extracted from the experimental data (see the main text). It is assumed that the  decoherence channels are the same for all four atomic levels.

The four-level atoms are held in an elongated MOT, which we model as a one-dimensional atomic medium of length $L$ with constant density along the $z$-axis.
Our simulations assume that the atomic medium does not modify the strong control beam with Rabi coupling strengths $\Omega_c^+(t)$ and $\Omega_c^-(t)$, whose propagation direction is orthogonal to the MOT axis; this approximation is validated by the experimental data. The Rabi coupling strength $\Omega_{\pi}(t)$, in contrast, can be modified by the atomic medium. To account for this, we introduce a spatially dependent Rabi coupling strength $\overline{\Omega}_{\pi}(z,t)$ [the bar is added to distinguish the spatially dependent Rabi coupling strength or spatially dependent field from the spatially-independent Rabi coupling strength $\Omega_{\pi}(t)$]. For $z=0$, i.e., at the ``entry point'' of the MOT, we have
\begin{eqnarray}
\label{eq_bc_omega}
    \overline{\Omega}_{\pi}(0,t)=\Omega_{\pi}(t).
\end{eqnarray}
The values of $\overline{\Omega}_{\pi}(z,t)$ for $z>0$ are determined by solving two groups of equations self-consistently.
The self-consistently determined quantity $|\overline{\Omega}_{\pi}(L,t)|^2$ is
proportional to the intensity $I_{\pi}$. It follows that $|\overline{\Omega}_{\pi}(L,t)|^2/|\Omega_{\pi}^{\text{max}}|^2$  is equal to $I_{\pi}/I_{\pi,0}^{\text{max}}$.  Theory results for $I_{\pi}/I_{\pi,0}^{\text{max}}$ are compared in Figs.~\ref{fig:slice_compare} and  \ref{fig:heatmap} of the main text
with experimentally measured intensity ratios.

Using Maxwell's equations, accounting for the medium polarizability, and making the slowly-varying envelope approximation, the propagation of the  probe pulse
 through the MOT can be modeled via~\cite{PhysRevLett.125.013601}
 \begin{eqnarray}
 \label{eq_maxwell}
 \frac{1}{c} \frac{\partial \overline{\Omega}_{\pi}(z,t)}{\partial t}
 + \frac{\partial \overline{\Omega}_{\pi}(z,t)}{\partial z} = - \imath \frac{\mu_a}{2} \sigma_{24}(z,t),
 \end{eqnarray}
where  $\mu_a$  is given by
\begin{eqnarray}
 \label{eq_mua}
  \mu_a = \frac{n_{\text{2D}} d^2 \omega}{\hbar \epsilon_0 c  L}.
\end{eqnarray}
Here,  $d$ and $\omega$ denote, respectively, the dipole moment and frequency of  the $D_2$ line, $\epsilon_0$ the free-space permittivity, and $c$ the speed of light. A naive estimate for the 2D density is given by $N/A$, where $A$ is the cross-sectional area of the MOT. Using this estimate, we find $c \mu_a=2.31 \times 10^6$~$\mu$s$^{-2}$. This estimate  does not take into account that the probe beam profile is smaller than the cross-sectional area of the MOT  and that the atom cloud is denser in the center than in the wings, nor does it take into account the optical depth. In terms of the peak optical depth $\text{OD}_{\text{peak}}$,
$c \mu_a$ can written as
\begin{eqnarray}
    \label{eq_mua}
    \mu_a =\frac{2 \text{OD}_{\text{peak}} \Gamma}{L}.
\end{eqnarray}
Using the measured peak optical depth of the imaging transition, we find
$c \mu_a=4.02 \times 10^7$~$\mu$s$^{-2}$. Since  estimating $c \mu_a$ accurately is challenging, we treat it as an adjustable parameter that is determined through detailed comparisons between simulation and  experimental data. The optimal value is found to be half of the  estimate given here (see Appendix~\ref{appendix:calibration} for details).

In Eq.~(\ref{eq_maxwell}), $\sigma_{24}(z,t)$ denotes a slowly varying dimensionless collective atomic operator.  The operators $\sigma_{jk}(z,t)$, where $j$ and $k$ both take the values $1$ to $4$, can be thought  of as representing elements of the atomic density matrix operator, averaged over a macroscopic number of atoms $N_z$, with $1 \ll N_z \ll N$. Their time evolution is governed by the following set of Heisenberg-Langevin equations:
\begin{widetext}
 \begin{subequations}
 \label{eq_atomic_operators}
    \begin{align}
        \frac{d\sigma_{11}(z,t)}{dt} =& \frac{\Gamma}{3}\sigma_{44}(z,t)
                           -\frac{\imath}{2}[\Omega_c^+(t)]^*\sigma_{14}(z,t)
                           +\frac{\imath}{2}\Omega_c^+(t)\sigma_{41}(z,t)
                         , \\
        \frac{d \sigma_{22}(z,t)}{dt} =& \frac{\Gamma}{3}\sigma_{44}(z,t)
                           -\frac{\imath}{2}[\overline{\Omega}_{\pi}(z,t)]^*\sigma_{24}(z,t)
                           +\frac{\imath}{2}\overline{\Omega}_{\pi}(z,t)\sigma_{42}(z,t)
                           , \\
        \frac{d \sigma_{33}(z,t)}{dt} =& \frac{\Gamma}{3}\sigma_{44}(z,t)
                           -\frac{\imath}{2}[\Omega_c^-(t)]^*\sigma_{34}(z,t)
                           +\frac{\imath}{2}\Omega_-(t)\sigma_{43}(z,t)
                           , \\
        \frac{d \sigma_{44}(z,t)}{dt} =& -\Gamma\sigma_{44}(z,t)
                           +\frac{\imath}{2}[\Omega_c^+(t)]^*\sigma_{14}(z,t)
                           +\frac{\imath}{2}[\overline{\Omega}_{\pi}(z,t)]^*\sigma_{24}(z,t)
                           +\frac{\imath}{2}[\Omega_c^-(t)]^*\sigma_{34}(z,t)
                           -\frac{\imath}{2}\Omega_c^+(t)\sigma_{41}(z,t)\nonumber \\
                         & -\frac{\imath}{2}\overline{\Omega}_{\pi}(z,t)\sigma_{42}(z,t)
                           -\frac{\imath}{2}\Omega_c^-(t)\sigma_{43}(z,t)
                           , \\
        \frac{d \sigma_{12}(z,t)}{dt} =& -\frac{\imath}{2}[\overline{\Omega}_{\pi}(z,t)]^*\sigma_{14}(z,t)
                           +\frac{\imath}{2}\Omega_c^+(t)\sigma_{42}(z,t)
                           -\imath E_1\sigma_{12}(z,t)
                           -\gamma_d{\sigma}_{12}(z,t)
                           , \\
        \frac{d \sigma_{13}(z,t)}{dt} =& -\frac{\imath}{2}[\Omega_c^-(t)]^*\sigma_{14}(z,t)
                           +\frac{\imath}{2}\Omega_c^+(t)\sigma_{43}(z,t)
                           +\imath (E_3 - E_1)\sigma_{13}(z,t)
                           -\gamma_d{\sigma}_{13}(z,t)
                           , \\
        \frac{d \sigma_{14}(z,t)}{dt} =& -\frac{\imath}{2}\Omega_c^+(t)\sigma_{11}(z,t)
                           -\frac{\imath}{2}\overline{\Omega}_{\pi}\sigma_{12}(z,t)
                           -\frac{\imath}{2}\Omega_c^-(t)\sigma_{13}(z,t)
                           +\imath (E_4 - E_1)\sigma_{14}(z,t)\nonumber \\
                         & +\frac{\imath}{2}\Omega_c^+(t)\sigma_{44}(z,t)
                           -\frac{\Gamma}{2}\sigma_{14}(z,t)
                           -\gamma_d{\sigma}_{14}(z,t)
                           , \\
        \frac{d \sigma_{23}(z,t)}{dt} =& -\frac{\imath}{2}[\Omega_c^-(t)]^*\sigma_{24}(z,t)
                           +\frac{\imath}{2}\overline{\Omega}_{\pi}(z,t)\sigma_{43}(z,t)
                           +\imath E_3\sigma_{23} (z,t)
                           -\gamma_d{\sigma}_{23}(z,t)
                           , \\
        \frac{d \sigma_{24}(z,t)}{dt} =& -\frac{\imath}{2}\Omega_c^+(t)\sigma_{21}(z,t)
                           -\frac{\imath}{2}\overline{\Omega}_{\pi}(z,t)\sigma_{22}(z,t)
                           -\frac{\imath}{2}\Omega_c^-(t)\sigma_{23}(z,t)
                           +\imath E_4\sigma_{24}(z,t)\nonumber \\
                         & +\frac{\imath}{2}\overline{\Omega}_{\pi}(z,t)\sigma_{44}(z,t)
                           -\frac{\Gamma}{2}\sigma_{24}(z,t)
                           -\gamma_d{\sigma}_{24}(z,t)
                           , \\
        \frac{d \sigma_{34}(z,t)}{dt} =& -\frac{\imath}{2}\Omega_c^+(t)\sigma_{31}(z,t)
                           -\frac{\imath}{2}\overline{\Omega}_{\pi}(z,t)\sigma_{32}(z,t)
                           -\frac{\imath}{2}\Omega_c^-(t)\sigma_{33}(z,t)
                           +\imath (E_4 - E_3)\sigma_{34}(z,t)\nonumber \\
                         & +\frac{\imath}{2}\Omega_c^-(t)\sigma_{44}(z,t)
                           -\frac{\Gamma}{2}\sigma_{34}(z,t)
                           -\gamma_d{\sigma}_{34}(z,t)
                           .
                           \label{eq_langevin}
    \end{align}
\end{subequations}
\end{widetext}
The equation for $\sigma_{kj}(z,t)$  with $k>j$ is the adjoint of the equation for $\sigma_{jk}(z,t)$. In writing Eqs.~(\ref{eq_atomic_operators}), we dropped $\delta$-correlated Langevin noise operators since they do not contribute at the level of approximation that we are working  at~\cite{PhysRevLett.84.4232,PhysRevA.65.022314}. All our calculations assume that the atoms are initially prepared in the $|F=1,m_F=0\rangle$ state, i.e., we seek self-consistent solutions to Eqs.~(\ref{eq_maxwell}) and
(\ref{eq_atomic_operators}) for the initial condition $\sigma_{jk}(z,t_{\text{ini}})=1$ for $j=k=2$ and $0$ otherwise. For each $(z,t)$ combination, we have $\sum_{j=1}^4 \sigma_{jj}(z,t)=1$.

\subsection{Numerical Approach}\label{secII}

For the numerical simulations, it is convenient to transform from the variables  $(z,t)$ to $(\xi,\Upsilon)$, where $\xi=z/c$ and $\Upsilon=t-z/c$; $\xi$ and $\Upsilon$ are variables in the reference frame of the probe pulse as it propagates through the medium. Note that $\xi$ and $\Upsilon$  both have units of time. Using the chain rule, one readily finds
\begin{eqnarray}
    \frac{1}{c} \frac{\partial}{\partial t} + \frac{\partial}{\partial z}= \frac{1}{c} \frac{\partial}{\partial \xi} .
\end{eqnarray}
Inserting this into Eq.~(\ref{eq_maxwell}), we have
\begin{eqnarray}
    \label{eq_maxwell2}
    \frac{\partial \overline{\Omega}_{\pi}(\xi,\Upsilon)}{\partial \xi}
    = - \imath \frac{1}{2} c \mu_a \sigma_{24}(\xi,\Upsilon).
\end{eqnarray}
The key advantage of Eq.~(\ref{eq_maxwell2}) over Eq.~(\ref{eq_maxwell}) is that the left hand side of Eq.~(\ref{eq_maxwell2}) can be evaluated using finite differencing in a single coordinate, namely the $\xi$-coordinate.  Since $z=0$ implies $\xi=0$ and $\Upsilon=t$,  the boundary condition from Eq.~(\ref{eq_bc_omega}) for the $(z,t)$ coordinates transforms to
\begin{eqnarray}
  \label{eq_bc_omega2}
  \overline{\Omega}(\xi=0,\Upsilon)=\Omega(t=\Upsilon)
\end{eqnarray}
for the $(\xi,\Upsilon)$ coordinates. To use $\overline{\Omega}_{\pi}(\xi,\Upsilon)$ in Eqs.~(\ref{eq_atomic_operators}), we need to transform Eqs.~(\ref{eq_atomic_operators}) from the $(z,t)$ coordinates to the $(\xi,\Upsilon)$ coordinates. For this, we note that $\partial/\partial t$ is equal to $\partial/\partial \Upsilon$. Moreover, for the ranges considered in this work, namely $z$ from $0$ to $3$~mm and $t$ from  $t_{\text{ini}}=-3$~$\mu$s to about $3$~$\mu$s, we see that $\Upsilon$ is to a very good approximation equal to $t$. In fact, the maximum error made by equating $\Upsilon$ and $t$ is around $2 \times 10^{-4}$~$\mu$s, which is smaller than the resolution of our time grid. We thus seek self-consistent solutions to Eqs.~(\ref{eq_atomic_operators}) [with $(z,t)$ replaced by $(\xi,\Upsilon)$] and (\ref{eq_maxwell2})  [with the boundary condition given in Eq.~(\ref{eq_bc_omega2})].

We pursue an iterative approach that rewrites Eq.~(\ref{eq_maxwell2}) using finite differencing:
\begin{eqnarray}
  \label{eq_maxwell3}
  \overline{\Omega}_{\pi}(\xi_l,\Upsilon) = - \imath c \mu_a \sigma_{24}(\xi_{l-1},\Upsilon) \Delta \xi + \overline{\Omega}_{\pi}(\xi_{l-1},\Upsilon). \nonumber \\
\end{eqnarray}
Here, $\xi_0,\xi_1\cdots,\xi_{M}$ are the grid points along the $\xi$-coordinate ($\xi_M=L/c$)  and $\Delta \xi$ is the spacing of the linear grid along $\xi$, $\Delta \xi=\xi_{l}-\xi_{l-1}$. Equation~(\ref{eq_maxwell3}) holds for $l=1,\dots,M$. The iterative scheme is as follows:
\begin{widetext}
  \begin{itemize}
    \item Step 0:
        \begin{enumerate}
            \item Set $\overline{\Omega}_{\pi}^{(0)}(\xi,\Upsilon) = \Omega(\Upsilon)$ for all $\xi$, and initialize $\sigma_{jk}^{(0)}(\xi,\Upsilon_{\text{ini}})=1$ for $j=k=2$ and $\sigma_{jk}^{(0)}(\xi,\Upsilon_{\text{ini}})=0$ otherwise.
            \item Solve Eqs.~\eqref{eq_atomic_operators} using $\overline{\Omega}_{\pi}(\xi,\Upsilon)=\overline{\Omega}_{\pi}^{(0)}(\xi,\Upsilon)$ to obtain $\sigma_{24}^{(0)}(\xi,\Upsilon)$.
       \end{enumerate}
   \item Steps $i = 1, \cdots$:
       \begin{enumerate}
        \setcounter{enumi}{2}
            \item Calculate $\overline{\Omega}_{\pi}^{(i)}(\xi_l,\Upsilon) = - \imath c \mu_a \sigma_{24}^{(i-1)}(\xi_{l-1},\Upsilon) \Delta \xi + \overline{\Omega}_{\pi}^{(i-1)}(\xi_{l-1},\Upsilon)$.
            \item Calculate $\overline{\Omega}_{\pi}^{(\text{new})}(\xi_l,\Upsilon)= y \overline{\Omega}_{\pi}^{(i)}(\xi_l,\Upsilon) +(1-y)\overline{\Omega}_{\pi}^{(i-1)}(\xi_l,\Upsilon)$, where $0 \le y < 1$.
            \item Initialize $\sigma_{jk}^{(\text{new})}(\xi,\Upsilon_{\text{ini}})=1$ for $j=k=2$ and $\sigma_{jk}^{(\text{new})}(\xi,\Upsilon_{\text{ini}})=0$ otherwise.
              \item Solve Eqs.~\eqref{eq_atomic_operators} using $\overline{\Omega}_{\pi}(\xi,\Upsilon)=\overline{\Omega}_{\pi}^{(\text{new})}(\xi,\Upsilon)$ to obtain $\sigma_{24}^{(\text{new})}(\xi,\Upsilon)$.
              \item If $\sum_{lm}\sum_{jk}|\sigma_{jk}^{(\text{new})}(\xi_l,\Upsilon_m)-\sigma_{jk}^{(i-1)}(\xi_l,\Upsilon_m)| < \epsilon$ ($\epsilon$ has a preset value), terminate the calculation. Otherwise set $\overline{\Omega}_\pi^{(i)}(\xi,\Upsilon)=\overline{\Omega}_\pi^{(\text{new})}(\xi,\Upsilon)$ and $\sigma_{jk}^{(i)}(\xi,\Upsilon)=\sigma_{jk}^{(\text{new})}(\xi,\Upsilon)$, increase $i$ by 1, and repeat the calculation beginning at step 3.
       \end{enumerate}
   \end{itemize}
\end{widetext}

The Heisenberg-Langevin equations are solved using a fourth-order Runge-Kutta algorithm with adjustable time step. Convergence of the collective atomic operators  is checked by systematically  decreasing the threshold value that governs the accuracy of the solutions. Moreover,  convergence is ensured by running simulations for different damping factors $y$ and convergence thresholds $\epsilon$. Due to the non-linearity of the equations, we occasionally find that the algorithm does not converge if $\epsilon$ is too small. Overall though, the outlined approach is quite robust.

\begin{figure}
    \centering
    \vspace*{0.1in} \includegraphics[scale=0.38]{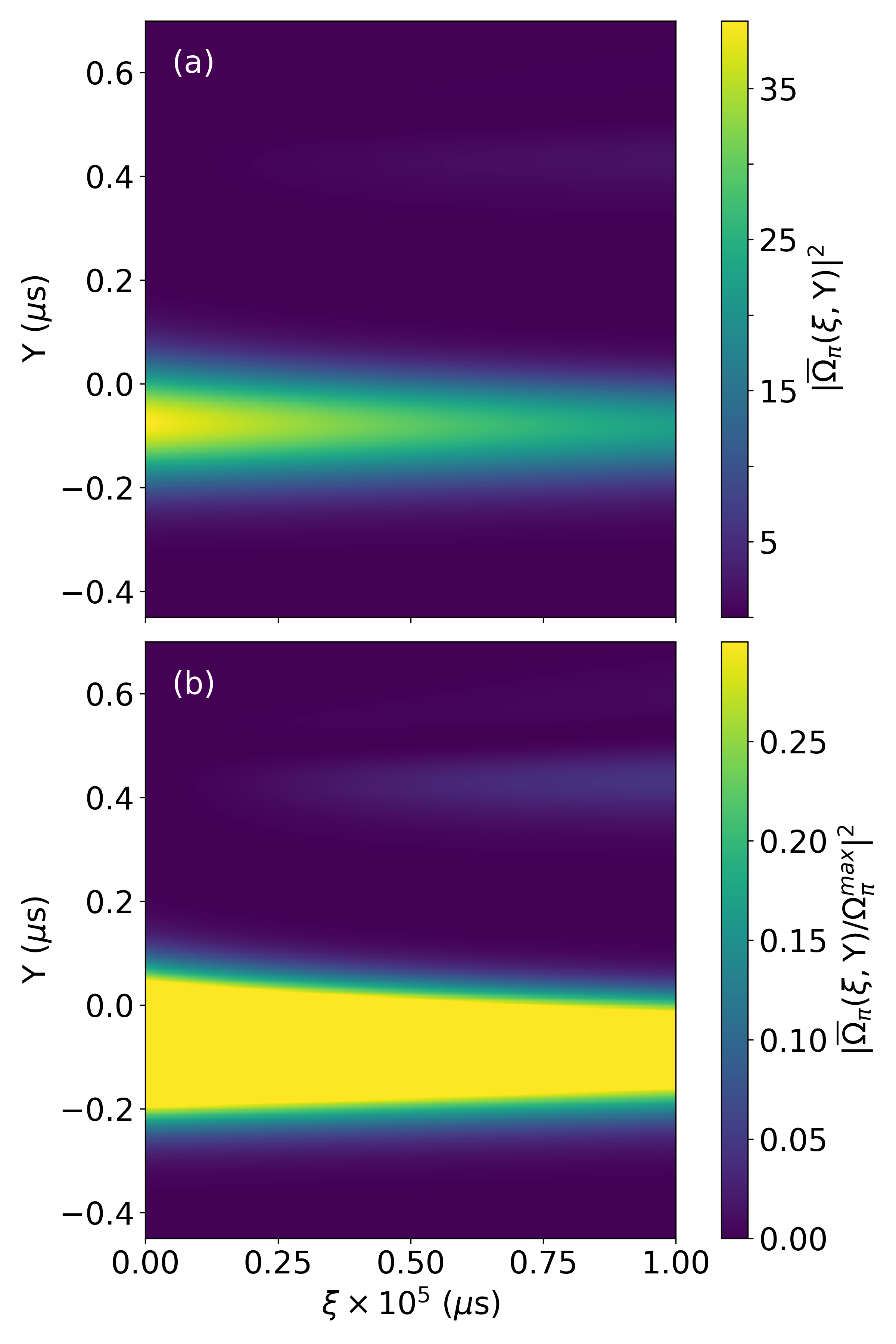}
    \caption{Modification of the probe beam by the atomic medium obtained using the algorithm outlined in Sec.~\ref{secII}. Panels~(a) and (b) show $|\overline{\Omega}_{\pi}(\xi,\Upsilon)|^2$ and $|\overline{\Omega}_{\pi}(\xi,\Upsilon)/\Omega_{\pi}^{\text{max}}|^2$, respectively, as functions of $\xi$ and $\Upsilon$. The simulation uses  $|E_1|=|E_3|=2 \pi \times 0.3  $~MHz, $\tau=0.4$~$\mu$s, $\Omega_c^{\max}=2\pi \times 9.5$~MHz, $\Omega_{\pi}^{\max}=2\pi$~MHz,  $c\mu_a=2.118\times10^7$~$\mu$s$^{-2}$,  $\chi=0$, and $\gamma_c=2\pi \times 0.111$~MHz.  The probe pulse is sent into the MOT at $\Upsilon=-0.07541$~$\mu$s (this is the scaled time at which the probe beam that enters the MOT is maximal). Initiated by the second control pulse (retrieval pulse), a weak  retrieval signal emerges around $\Upsilon=0.4$~$\mu$s for $\xi>0$.}
    \label{probe}
\end{figure}

Figure~\ref{probe}(a) shows the result of one of our simulations. Specifically,
Fig.~\ref{probe}(a) shows $|\overline{\Omega}_{\pi}(\xi,\Upsilon)|^2$ as functions of $\xi$ and $\Upsilon$ for $\Omega_c^{\max}=2 \pi \times 9.5  $~MHz, $\Omega_{\pi}^{\max}=2 \pi$~MHz, $\gamma_c=2 \pi \times 0.111  $~MHz, $c\mu_a=2.118\times10^7$~$\mu$s$^{-2}$, $\tau=0.4$~$\mu$s, $\chi=0$, and $|E_1|=|E_3|=2 \pi \times 0.3$~MHz.
The values of $\Omega_c^{\max}$, $\Omega_{\pi}^{\max}$, and $\gamma_c$ are the same as those used in  the simulations shown in Figs.~\ref{fig:slice_compare} and \ref{fig:heatmap} of the main text. It can be seen that $|\overline{\Omega}_{\pi}(\xi,\Upsilon)|^2$ decreases, for $\Upsilon \approx -0.075$~$\mu$s, with increasing $\xi$ [this is the $\Upsilon$ value for which $|\overline{\Omega}_{\pi}(0,\Upsilon)|^2$ is maximal]. The decrease is accompanied by a ``new signal'' (i.e., the ``retrieved signal''), which emerges with increasing $\xi$ at $\Upsilon \approx 0.4$~$\mu$s. To highlight the retrieved signal, Fig.~\ref{probe}(b) shows the ratio $|\overline{\Omega}_{\pi}(\xi,\Upsilon)/\Omega_{\pi}^{\text{max}}|^2$, with the color bar capped at $0.3$. In this representation, the retrieved signal is clearly visible.

\subsection{``Calibration'' of $\Omega_{c}^{\text{max}}$,  $\Omega_{\pi}^{\text{max}}$, and $\mu_a$} \label{appendix:calibration}

Equations~(\ref{eq_maxwell})-(\ref{eq_langevin}) contain two parameters ($\Omega_{c}^{\max}$ and $\Omega_{\pi}^{\max}$) that  cannot, with our  set-up, be determined through independent experimental measurements. Instead, we determine the values  by seeking good overall agreement between theory and experiment.   Moreover, even though $\mu_a$ can, as discussed in Appendix~\ref{appendix:theoryframework} be  estimated roughly from experimentally measured quantities, those estimates have some uncertainties. We thus use comparisons between simulation results and experimental data to determine the ``optimal'' value of $\mu_a$. As discussed below, we performed a good number of scans to determine the optimal values of $\Omega_{c}^{\max}$, $\Omega_{\pi}^{\max}$, and $\mu_a$. Importantly, the simulations shown in Figs.~\ref{fig:slice_compare} and \ref{fig:heatmap} of the main text  all  use the same values for $\Omega_{\pi}^{\text{max}}$, $\Omega_{c}^{\text{max}}$, and $\mu_a$.

\begin{figure}
    \vspace*{0.2in}\centering
    \includegraphics[scale=0.35]{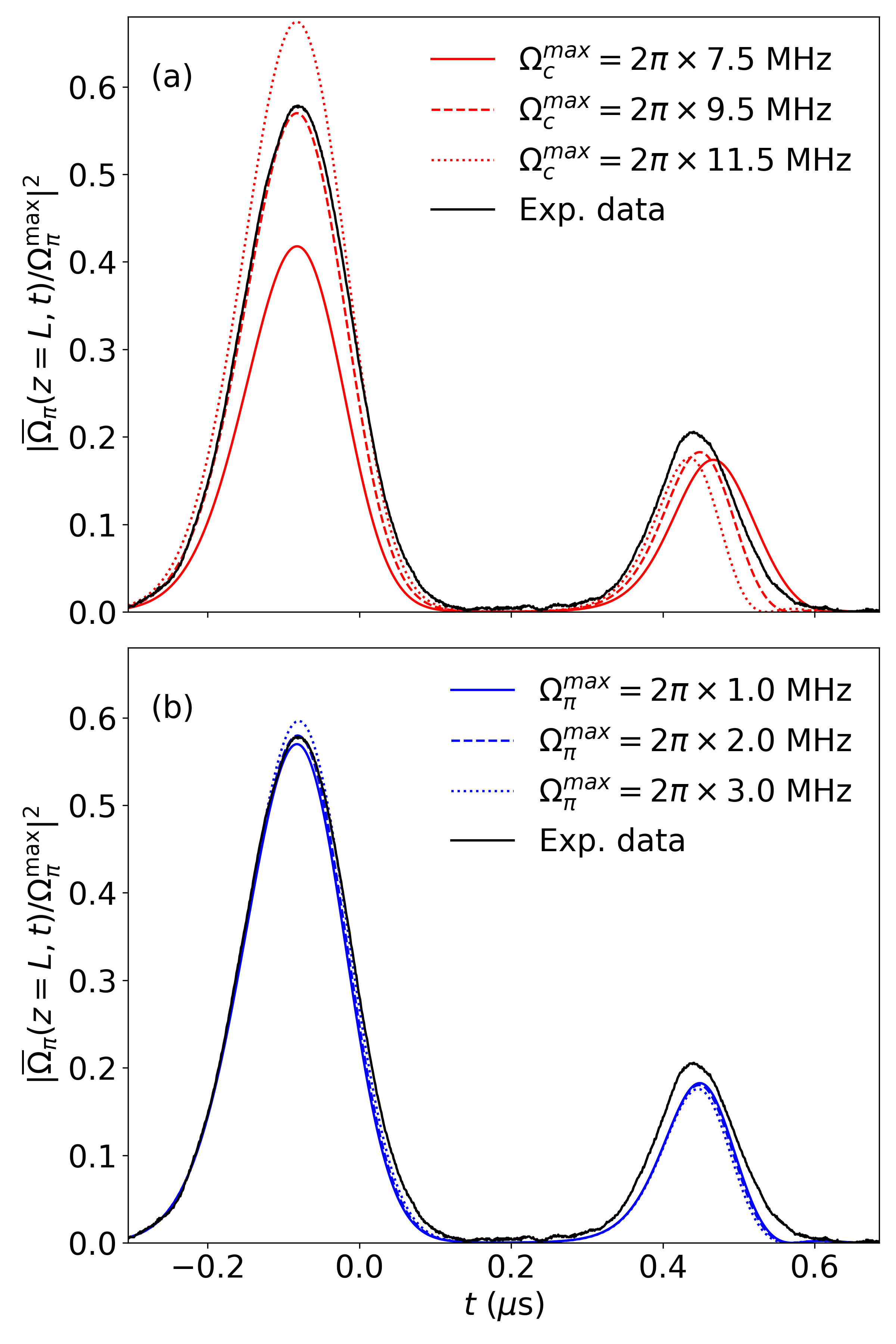}
    \caption{Normalized probe intensity  $|\overline{\Omega}_\pi(z=L,t)/\Omega_{\pi}^{\text{max}}|^2$ for $B=\chi=0$, $\tau=0.4$~$\mu$s, $\gamma_c=2\pi \times 0.111$~MHz, and $c\mu_a=2.118\times10^7$~$\mu$s$^{-2}$. In~(a), the simulations fix the maximal probe beam coupling  at $\Omega_\pi^\text{max}=2\pi$ MHz while the maximal control beam coupling $\Omega_c^\text{max}$ is varied (red lines, see legend). In~(b), the simulations fix the maximal control beam coupling  at  $\Omega_c^\text{max}=2\pi \times 9.5$ MHz while $\Omega_\pi^\text{max}$ is varied (blue lines, see legend). For comparison, experimental data are shown in both panels by black lines.}
    \label{fig:scan_rabi}
\end{figure}

Figures~\ref{fig:scan_rabi}(a) and \ref{fig:scan_rabi}(b) illustrate how  $|\Omega_{\pi}(z=L,t)|^2/|\Omega_{\pi}^{\text{max}}|^2$ changes when
$\Omega_c^{\max}$ and  $\Omega_{\pi}^{\max}$, respectively, are changed for otherwise constant parameters (see figure caption for details). Figure~\ref{fig:scan_rabi}(a) shows that the peak height of the normalized transmitted  probe signal and the peak position of the retrieved signal depend, for fixed $\Omega_{\pi}^\text{max}$,  appreciably on $\Omega_{c}^{\text{max}}$.  Specifically, the height of the transmitted probe peak increases with increasing $\Omega_c^\text{max}$ while the position of the retrieved probe shifts toward smaller times with increasing $\Omega_c^\text{max}$. Figure~\ref{fig:scan_rabi}(a) shows that $\Omega_c^{\text{max}}=2\pi \times 9.5$~MHz provides the best agreement with the experimental data.   Figure~\ref{fig:scan_rabi}(b) shows that the transmitted and retrieved probe depend, as expected, comparatively weakly on  $\Omega_{\pi}^\text{max}$  in the regime where $\Omega_{\pi}^\text{max} \ll \Omega_c^\text{max}$. Comparing theory and experiment for a good number of scans, we choose $\Omega_c^\text{max}=2\pi \times 9.5$ MHz and $\Omega_{\pi}^\text{max}=2\pi$ MHz for the simulations presented in the main text.

\begin{figure}
    \vspace*{0.2in}\centering
    \includegraphics[scale=0.35]{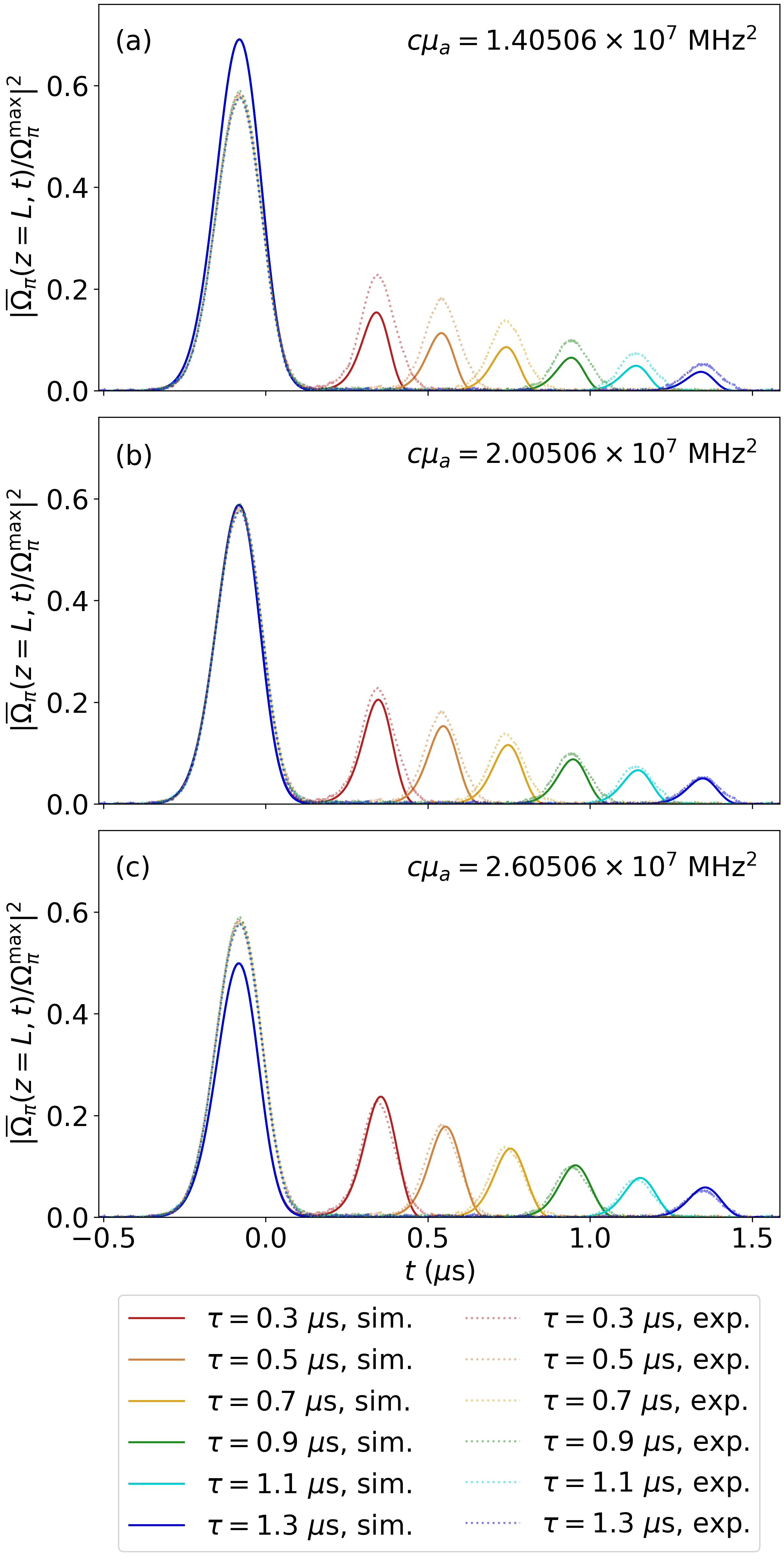}
    \caption{Normalized probe intensity $|\overline{\Omega}_{\pi}(z=L,t)/\Omega_{\pi}^{\max}|^2$ for $B=\chi=0$ as a function of $t$ for various delay times $\tau$ [see the legend below panel~(c)]. Simulation results are shown by bright colors and solid lines while experimental data are shown by faint colors and dotted lines. The simulations use $\Omega_c^{\text{max}}=2\pi \times 9.5$~MHz, $\Omega_{\pi}^{\text{max}}=2\pi$~MHz, $B=\chi=0$, and $\gamma_c=2\pi \times 0.111~\text{MHz}$. While the experimental data are the same for panels~(a)-(c), the simulations use different $c\mu_a$ (see legend).
    The best agreement between experiment and theory is obtained for $c\mu_a=2.118\times10^7$~MHz$^{2}$. }
    \label{fig:scan_mua}
\end{figure}

Last, we want to understand how the effective medium-laser coupling  parameter $c\mu_a$ impacts the probe field. Figures~\ref{fig:scan_mua}(a)-\ref{fig:scan_mua}(c) shows that the height of the transmitted probe peak decreases with increasing $c \mu_a$ while  the height of the retrieved peaks increases with increasing $c \mu_a$. The peak positions are found to depend comparatively weakly on the value of $c \mu_a$, at least for the parameter combinations considered.
Based on the results shown in Fig.~\ref{fig:scan_mua}, we chose $c\mu_a=2.118\times10^7~\text{MHz}^2$, which is about half as large as  the  $\text{OD}_{\text{peak}}$-based estimate from Section~\ref{appendix:theoryframework}. Note that the digits after the decimal point are not necessarily meaningful; they are reported for reproducibility purposes.

\end{document}